# Next Frontiers in Particle Physics Detectors: INSTR2020 Summary and a Look into the Future

**Maxim Titov**[a]

[a] *Commissariat à l'Énergie Atomique et Énergies Alternatives (CEA) Saclay, DRF / IRFU / DPHP, 91191 Gif sur Yvette Cedex, France*
  *E-mail*: maxim.titov@cea.fr

ABSTRACT: The physics goals of high luminosity particle accelerators, from LHC to HL-LHC and to the next generation of lepton colliders, have set quite stringent constraints on the future needs at the Instrumentation Frontier. Many technologies are reaching their sensitivity limit and new approaches need to be developed to overcome the currently irreducible technological challenges. The detrimental effect of the material budget and power consumption represents a very serious concern for a high-precision silicon vertex and tracking detectors. One of the most promising areas is CMOS sensors offering low mass and potentially radiation-hard technology for the future proton-proton and electron-positron colliders, intensity frontier and heavy-ion experiments. MPGDs have become a well-established technique in the fertile field of gaseous detectors; these will remain the primary choice whenever the large-area coverage with low material budget is required. Vacuum tube technology is inherently fast and new developments include advances in microchannel plates for photomultipliers with a potential for a picosecond-time resolution in large systems. Several novel concepts of picosecond-timing detectors will have numerous powerful applications in particle identification, pile-up rejection and event reconstruction, and serve numerous scientific goals. The story of modern calorimetry is a textbook example of physics research driving the development of an experimental method. Silicon photomultipliers have seen a rapid progress in the last decade, becoming the standard solution for scintillator-based devices. The integration of advanced electronics and data transmission functionalities plays an increasingly important role and needs to be addressed. Bringing the modern algorithmic advances from the field of machine learning from offline applications to online operations and trigger systems is another major challenge. The timescales spanned by future projects in particle physics, ranging from few years to many decades, constitute a challenge in itself, in addition to the complexity and diversity of the required accelerator and detector R&D. This paper summarizes advances and recent trends in the instrumentation techniques for particle physics experiments, largely based on the presentations given at the International Conference "Instrumentation for Colliding Beam Physics" (INSTR-20), held at BINP Novosibirsk, Russia, from 24 to 28 February, 2020.

**Contents**

**1. Introduction**

Today, particle detectors are key to address future science challenges and their development is based on our understanding of fundamental laws of physics. Our instrumentation represents both a towering achievement, and, in some cases, a scaled-up version of techniques used in the past. Recent discoveries of the Higgs boson and Gravitational Waves required increasingly sophisticated detectors and have created an exceptionally positive environment in society. Thus we have a "virtuous cycle" which must remain strong and un-broken – laws of nature enable novel detector concepts and techniques, which in turn lead to a greater physical discoveries and better understanding of our Universe. Advanced innovative accelerator concepts able to push the present limits of energy and luminosity at future frontier colliders. In this context, detectors in high-energy physics face a huge variety of operating conditions and employ technologies that are often deeply entwined with developments in industry. The environmental credentials of detectors are also increasingly in the spotlight. Society is the principal client, and many of the accelerator innovations and particle physics technologies are vital for other applications, such as IT communications, biology and medical imaging, and health.

At the Energy Frontier, one can distinguish two major drivers for detector R&D in a short-term: detector upgrades towards HL-LHC and development of advanced technologies for the future e+e- colliders. The latter are inherently very accurate physics probes requiring integrated concepts with ultimate precision, minimal power consumption and ultra-light structures. A key challenge to demonstrate detector performance needed in a realistic environment represents a major step forward towards a complete system. Fast and accurate physics-driven detector simulation have gained considerable importance with the increase of the complexity of novel instrumentation. All detector and electronics developments need test beam and irradiation facilities to prove the performances and to verify radiation effects on active parts and passive materials. Future experimental projects also face a large number of diverse engineering challenges, in the areas of system integration, power distribution, cooling, mechanical support structures, and production techniques. The use of advanced machine learning algorithms, such as neural networks (NN), boosted decision trees (BDT) and many others, is a long-standing tradition in particle physics since 1990's [1] and has been already key enabler for discoveries (e.g. single-top production at Tevatron [2]). These techniques are expected to grow to address a number of challenges that arises from both the computational and physics environment of the next-generation experiments. In addition, offline-like reconstruction is increasingly possible in online triggering systems, with further reduction in the offline data volume without loss of physics performance. Within the field of particle physics, technologies developed under generic R&D studies or with the aim to address experiment-oriented challenges at future colliders provide a boost in innovation and novel designs that often suit the needs of the Intensity or Cosmic Frontiers, i.e. neutrino or astroparticle physics.

## 2. Large Scale Science Projects in Particle Physics

Large-scale science projects (in particle physics) at the forefront of research and technology require large and sustained infrastructures, of which big laboratories and global collaborations on long-term scale are the key constituents. The mission of research infrastructures is driven by four elements: research, innovation, education and outreach. The open collaborative approach with competent partners provides the opportunity for people to learn working together, it serves as a forum for co-operation and competition, encouraging the sharing of information and promoting diversity through the recognition of differences [3]. All this can be seen as a model for future cooperation in the world in many fields beyond science, representing one of the most important benefits of particle physics to society. During more than 50 years, CERN served as a cornerstone of modern research infrastructure as well as the organizational and legal framework and can proudly look back on a long history of successful realizations of ever increasingly complex accelerators (PS, ISR, SPS, LEP, LHC). In addition, a number of particle physics laboratories have been built across the globe, operating at a regional and/or national levels, and creating a solid basis of R&D platforms that were crucial for the advances of the high energy physics.

Almost all large-scale facilities have been constructed with some level of in-kind contribution from other regions across the globe. One of the European examples is DESY/Hamburg laboratory, who was the host of DORIS, PETRA and HERA accelerators from 1959 and nowadays operates the European X-Ray Free Electron Laser (XFEL) facility. The latter is based on superconducting RF-cavities (SRF) and serves as an ultimate integrated test (10%) and the "technology base" for the future International Linear Collider (ILC). The largest LNF/INFN laboratory in Italy, founded in 1954, is still centered on the DAFNE e+e- accelerator complex, but in the recent years, a second infrastructure, SPARC-LAB, has been developed to study new techniques for particle acceleration. Many interesting ideas can be exploited at LNF in the future, including technological tests for muon colliders (LEMMA), study of electron cloud effects for HL-LHC and FCC, searches for dark photon [4], and plasma cells operated laser-driven/particle-driven plasma wake-field acceleration (LWFA/PWFA) mode to accelerate electrons up to 1-2 GeV [5]. The heavy ion collisions at FAIR (Darmstadt, Germany) and NICA (JINR, Dubna,

Russia) permit the exploration of the "terra incognita" of the QCD phase diagram in the region of high baryon densities. Being an ESFRI "Landmark Project" similar to the HL-LHC [6], the FAIR construction plan foresees a staged completion of the facility that would allow first experimental programs to commence as early as 2024 (Phase I), while the entire facility would be completed in 2026-2027 (Phase III) [7]. The United States has a rich history of particle physics accelerator facilities at BNL, SLAC, JLAB and Fermilab. The LBNF/DUNE program [8], based on the Fermilab's first MW beam of protons, will be the first internationally conceived, constructed, and operated megascience facility in the US, while BNL will serve as a host for the Electron-Ion Collider (EIC) – the highest priority for the near future in the US nuclear physics. In Asia, the IHEP CAS manages or hosts a number of current (and future) China's major science facilities in space missions, underground neutrino physics, and accelerator-based ones, from Upgraded Beijing Electron–Positron Collider (BEPC) to the future Circular Electron–Positron Collider (CEPC) [9]. A new 4500 $m^2$ SRF Frontier R&D laboratory was constructed in Huairou Science Park, Beijing, with a potential to contribute to the SRF cavity and cryomodule mass-production. Established in 1971, KEK in Japan serves today as the host of the upgraded to the SuperKEKB collider in Tsukuba aiming for a leap of a 40 times the KEKB luminosity and operates the upgraded JPARC accelerator complex, which will provide a high-intensity neutrino beam to the Hyper-Kamiokande (HyperK) to explore CP-violation in the neutrino sector [10]. The potential of HyperK experiment, which was approved in Feb. 2020 and has an 8.4 times larger fiducial mass than its predecessor, is similar to the Fermilab LBNF/DUNE project, but it uses different detector technology (water Cherenkov detectors).

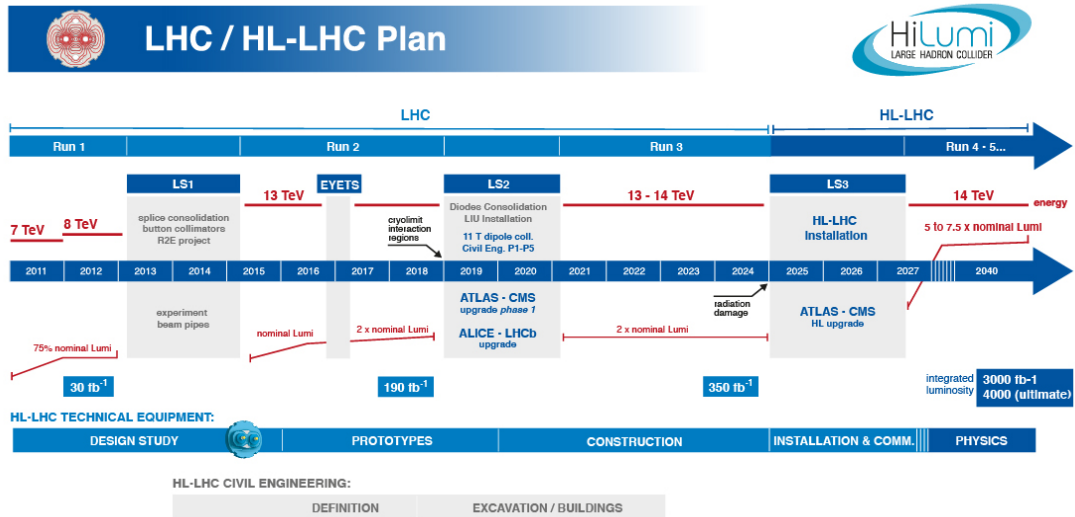

**Figure 1.** Timeline of the LHC baseline programme and its upgrade phases, showing the energy of the collisions (upper line - red) and integrated luminosity (lower lines - blue).

The pursuit of ever-higher energies will surely be one of the future directions of elementary particle physics; the course will depend on whether one can continue to contain the cost of the future colliders. Over the past six decades, the center-of-mass energy has been increased by five orders of magnitude and the luminosity by about seven orders [11]. The LHC is an outstanding flagship of the collider physics and a culmination of this evolution. In the LHC Run 2 (from 2015 through 2018), the ATLAS and CMS experiments collected about 160 fb$^{-1}$ and the LHCb experiment about 10 fb$^{-1}$ of data in pp-collisions at the centre-of-mass energy of 13 TeV, surpassing by a factor of two the design value of ATLAS/CMS instantaneous luminosity of $10^{34}$

cm$^{-2}$s$^{-1}$ [12]. Detailed studies are being carried out of the Higgs boson and Standard Model (SM) processes as well as searches for physics beyond the SM, including heavy flavour sector. The major questions in particle physics today are: what could be the "Nature of the New Physics" and what is the "Next Energy Physics Scale" for building a new collider.

The LHC is intended to accumulate some 350 fb$^{-1}$ by the end of Run 3 (~2024), as shown in Figure 1. After the third long shutdown (LS3), the HL-LHC operational phase is scheduled to commence in 2027. It will rely on a number of key innovative technologies, including cutting-edge 11-12 T superconducting magnets, compact superconducting crab cavities with ultra-precise phase control for beam rotation, new technology for beam collimation, high-power, loss-less superconducting links, and various components to cope with an increased radiation and stored energy. The high-luminosity upgrade of the LHC will increase the proton–proton collision dataset by an order of magnitude to 3000 fb$^{-1}$ over the period of operation in 2027–2037. It is also planned to "level" the instantaneous luminosity at about ~ $5 - 7.5 \times 10^{34}$ cm$^{-2}$ s$^{-1}$ during HL-LHC. The full exploitation of the LHC, including its upgraded version, the HL-LHC, was defined as the top priority in the 2013 European Strategy Update for Particle Physics [13].

The Lepton Colliders currently operating or planned are DAFNE at LNF/INFN, Italy, BEPC II and CEPC in China, SupekKEKB and ILC in Japan, Compact Linear Collider (CLIC) and Future Circular Collider (FCC-ee) at CERN, the Super Charm-Tau factory (SCT) at BINP, Novosibirsk, and the High Intensity Electron Positron Accelerator (HIEPA) in China. Despite the fact that all these facilities have different energy ranges (from 0.5 GeV per beam at DAFNE, up to 1 TeV at ILC and 3 TeV final energy at CLIC), different sizes (from 100 m for DAFNE up to 100 km for FCC-ee), different configurations (linear versus circular) and luminosities differing by orders of magnitude (from $10^{34}$ to $10^{36}$ cm$^{-2}$ s$^{-1}$), there are many similarities in the collider designs. A novel "Crab Waist Collision Scheme", proposed in 2006 and implemented at DAFNE, requires very low emittance beams, and is of particular importance for the SCT and FCC-ee colliders; the latter requires vertical emittance less or equal than 1 pm at 45 GeV. The BEPC-II at IHEP (Beijing, China) and VEPP-4M / VEPP2000 in Novosibirsk [14] with 3 detectors (KEDR, SND and CMD-3) are in operation for more than 20 years, and went through a long series of optimizations and upgrades. Unique to VEPP-4M is the precision determination of the beam energy, using the resonant depolarization method, with an ultimate accuracy of $10^{-6}$. Accurate energy calibration using this concept or laser Compton backscattering technique [15] is essential for the ILC [16] or FCC-ee [17] (FCC-ee aims for Z-boson mass measurement with a precision of 2x10$^{-6}$). Future SCT [18] and HIEPA factories are designed as a low energy electron-positron colliders with high luminosity of ~ 1-2 x 10$^{35}$ cm$^{-2}$s$^{-1}$ at a centre-of-mass energy between 2 GeV and 7 GeV with a possibility to exploit longitudinally polarized electrons at the interaction point. The SuperKEKB machine, with a design peak luminosity of 8×10$^{35}$ cm$^{-2}$s$^{-1}$ and a target integrated luminosity of 50 ab$^{-1}$, was able to achieve luminosities above 1.5x10$^{34}$ cm$^{-2}$s$^{-1}$ early in 2020, similar to KEKB, with a modest beam current of ~500mA. This is a clear success of colliding bunches with a very small spot sizes achieved by the low-emittance and nanometer-scale beams [19]. Higher luminosity in SuperKEKB is expected after enough beam-scrubbing, and studies of the Crab Waist concept would still be required in the next run. One of the critical parts of Crab Waist colliders is their final-focus arrangement system. Another common study topic for the SuperKEKB and FCC-ee is related to the high total beam current effects (1-2 A), including beams causing dangerous collective instabilities, ion clouds (in the electron ring) and electron clouds (in the positron ring), as well as the heating of vacuum components by the RF bunch field.

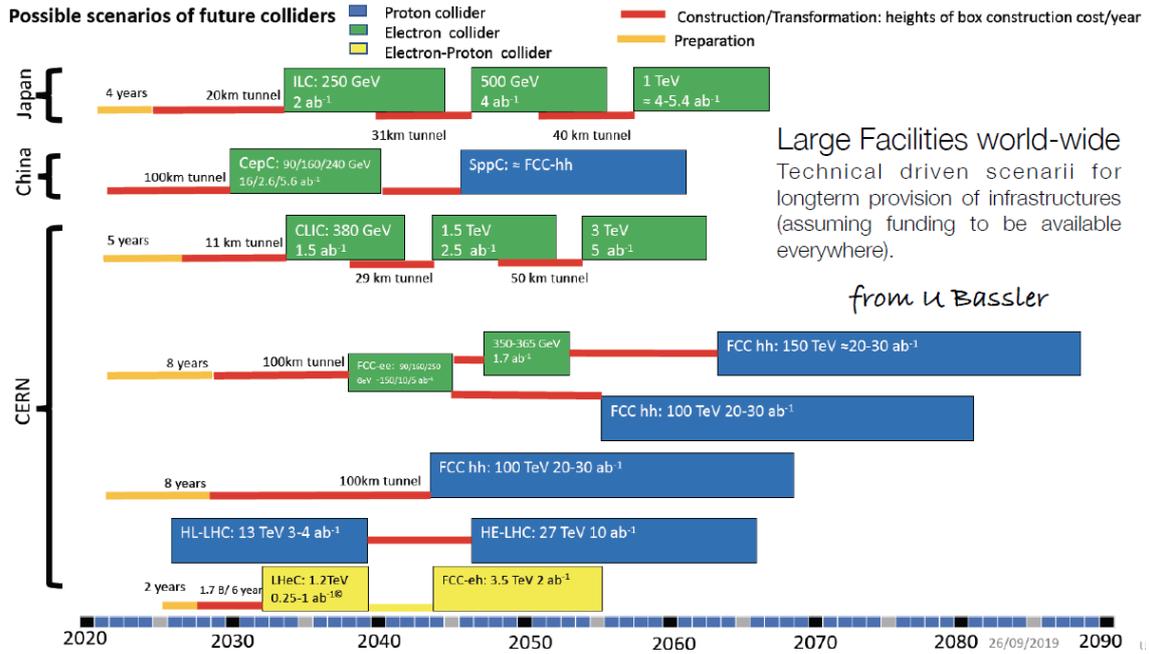

**Figure 2.** Approximate technically limited timelines of Large Future Colliders, as extracted from inputs submitted to the 2018 European Strategy Update (by U. Bassler) [12].

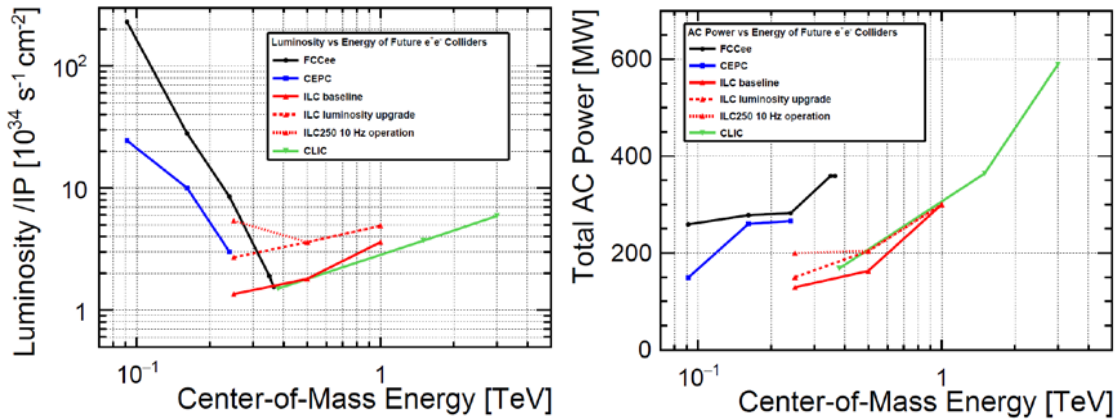

**Figure 3.** (left) Luminosity of the proposed Higgs/Electroweak factories (CePC, CLIC, FCC-ee, ILC) as a function of the center-of-mass energy. The numbers are given per interaction point (IP), while the effect of polarization is not included for Linear Colliders. The CDRs assume 2 IP for FCC-ee and CepC. ILC luminosity upgrades are possible by doubling the number of bunches per pulse and by an increase in the pulse repetition rate from 5 to 10 Hz; (right) Site AC power requirements for future e+e- colliders [24].

In the discussions during the Open Symposium on the European Particle Physics Strategy Update (EPPSU) (May 13-16, 2019, Granada, Spain) [20], a "Higgs Factory" has been considered as the highest scientific priority among all accelerator options after CERN's LHC. Due to the Higgs mass of 125 GeV, four proposals been put forth: the 250 GeV ILC in Japan, based on SRF-cavities, the 380 GeV CLIC at CERN, based on a novel two-beam acceleration technique, both colliders upgradable to O(TeV) energy range, and the two circular options, FCC-ee at CERN and the CEPC in China both utilizing ~O(100 km) tunnel. The FCC-ee and CEPC are proposed to be followed by a proton–proton colliders, FCC-hh and SppC, respectively.

Among all projects, the ILC (250 GeV), as a Higgs Factory in its first stage and upgrade provision for 1 TeV, is by far the most advanced in terms of technology, maturity, cost, and preparations in international cooperation [21]. It is based on the Technical Design Report and ready for construction. In its statement early this year [22], the International Committee on Future Accelerators (ICFA) advocates establishment of an international development team to facilitate transition into the ILC pre-lab phase by summer 2021. At this point it would be possible to launch the inter-governmental negotiations, provided Japan expresses the intent to do so together with international partners, before the establishment of the ILC Laboratory for construction (in about 4 years). Other projects (CLIC, CePC, FCC-ee) are currently exist as Conceptual Design Reports (CDR). Figure 2 illustrates approximate technically limited timelines of future large-scale collider facilities for the next five decades, based on the presentations by their proponents given in Granada, and summarized in the Physics Briefing Book [23]. Figure 3 (left) presents luminosity of the proposed Higgs factories as a function of center-of-mass energy [24]. Energy efficiency is one of several important factors in making high-energy physics more sustainable in the long term. Figure 3b (right) shows the site AC power requirements as a function of collision energy.

For the future Energy Frontier proton-proton colliders, three current options - CERN's HE-LHC (27 TeV), FCC-hh (100 TeV) and SppC (80 TeV) - demand high-field superconducting dipole magnets of 16-20 T, based on $Nb_3Sn$ or HTS technologies. There are natural constraints for such an advanced-magnet development regardless of budget and manpower involved. Several facilities for neutrino physics, based on high power proton beam accelerators, are operational or being developed. One of the possible outcomes of these facilities is a future muon collider. Its advantages, apart from direct Higgs production in s-channel, are compactness, low synchrotron-radiation power, small energy spread at the interaction point, and negligible beamstrahlung.

Particle physics is now entering a new era. As the scale and the cost of the frontier machines increases, while the timescale for projects is becoming longer, fewer facilities can be realized. Moreover, several high-energy physics laboratories becoming multi-purpose ones. To maintain a global vision of our domain going beyond the regional/national boundaries, it is important to maintain particle physics laboratories and accelerator expertise in all regions. A pure centralization does not solve the current uncertainty for the future of high energy physics, but will lead to a decrease in overall financial resources. The next-generation of Energy Frontier colliders are all multi-billion projects, which can only be realized with more collaboration and coordination on a global scale. The particle physics community will need to define the most appropriate governance structure for the next (post-LHC) facility, independent of its location, where CERN should be the hub of European activities, providing coordination and support at different levels. Our culture and management structure must evolve to confront these challenges.

**2.1 Future Challenges: Generic Detector R&D, Computing and Software**

Emerging novel detector technologies are the vital backbone for the success of the upcoming large and complex particle physics experiments. They are usually the result of a long development cycle encompassing many stages: generic "blue-sky" R&D and R&D activities guided by the needs of future projects, focused R&D activities for an approved experiment, production, industrialization and installation/commissioning, where engineering aspects play a very important role. On average, it might take more than 20 years to mature a technology from the original idea to a well-established technique suited for implementation (e.g. the first workshop on the LHC was held in 1984). In the 1990s, there was a large-scale R&D program, monitored by the Detector

R&D Committee (DRDC) at CERN [25]. Such a "cell-oriented" approach (DRDC and later CERN White Paper R&D programme in 2008-2011) were crucial to streamline effort/resources, handle new techniques and common components to on-going detector engineering, and finally to select particular technologies for design and construction of LHC detectors and later for HL-LHC upgrades. The long timescale of particle physics projects and the complexity of the required instrumentation drive the need for strong support for technical infrastructure at the regional/national laboratories and larger institutions, working in conjunction with the global one, like CERN.

High energy physics (HEP) driven R&D has always played a prominent role in providing cutting-edge technologies. Looking ahead, there is a clear need to strengthen existing R&D collaborative structures and create new ones, to address future detector, electronics and computing challenges after HL-LHC. Consistent with this view, CERN has developed a strategic R&D programme [26] that concentrates on advancing key technologies rather than on developing specialized detector applications. In addition, many novel developments are being carried out exploiting existing CERN-RD collaboration structures, such as RD42 [27], RD50 [28], RD51 [29], RD53 [30], and with industrial partners. The main topics of the RD50 research program focus on understanding of displacement damage and its impact on macroscopic properties of silicon detectors, irradiation tests campaigns and providing new structure development. The RD51 collaboration and MPGD success is related to the RD51 model in performing R&D: combination of generic and focused R&D with bottom-up decision processes, full sharing of "know-how", information, common infrastructures. This model has to be continued and can be exported to other detector domains. The RD51 structure with seven working groups is illustrated on Figure 4. One of the tasks of CERN-RD53 collaboration is to design novel architecture - new readout chips for the ATLAS and CMS pixel systems at HL-LHC that can take advantage of a more downscaled 65 nm CMOS process.

The landscape of proposed energy frontier accelerator facilities is broad in terms of detector technologies to address the physics programmes [31]. Concerning synergies and challenges in the design and instrumentation developments for the ILC and CLIC (and future circular lepton collider) experiments, these are in most cases orthogonal to the main directions of HL-LHC experiments. Rather than emphasizing radiation hardness and rate capability, the demands for resolution (granularity) and material budget on one hand, and on acceptable speed and power consumption on the other hand, exceed significantly what is the state-of-the-art today. Such leaps in performance cannot be achieved by simple extrapolation of the known, but only by entering new technological territory in detector R&D. Several new concepts for silicon sensor integration, such as monolithic devices, are being pursued for pixel vertex detectors, new micro-pattern gas amplification techniques are explored for tracking and muon systems, and the particle-flow approach to calorimetry promises to deliver unprecedented jet energy resolution, to quote just some examples.

Over the past few decades, a large number of groups have pursued extensive generic detector R&D studies [32] applicable to any of the proposed Linear Collider detector concepts (International Large Detector (ILD) [33], Silicon Detector (SiD) [34] and CLIC Detector and Physics (CLICdp) [35]. Intentionally, these groups did not yet make very specific choices and keep various options for technologies to realise the individual sub-detectors. This has the advantage that the technologies can be further developed until specific choices have to made once the project is approved. Furthermore - and as important - this keeps a broad community of detector research groups at universities and laboratories involved and increases the chance to arrive at the best technically possible detector solution when it has to be built. Wherever possible, the R&D is not pursued with respect to one specific concept but in the context of world-wide transversal R&D collaborations which work on technologies rather than detector designs, such as CALICE [36], LCTPC [37], and FCAL [38]. Further R&D programmes are ongoing, e.g. for various monolithic pixel detector technologies (CMOS, DEPFET, FPCCD, SoI), for generic aspects of the next generation silicon vertex detectors for Linear Colliders and for the development of forward calorimetry. Last, but not least, European Commission funded programmes, such as EUDET [41], AIDA/AIDA2020 [42] and ATTRACT [43] play an important role in enabling and supporting generic R&D activities.

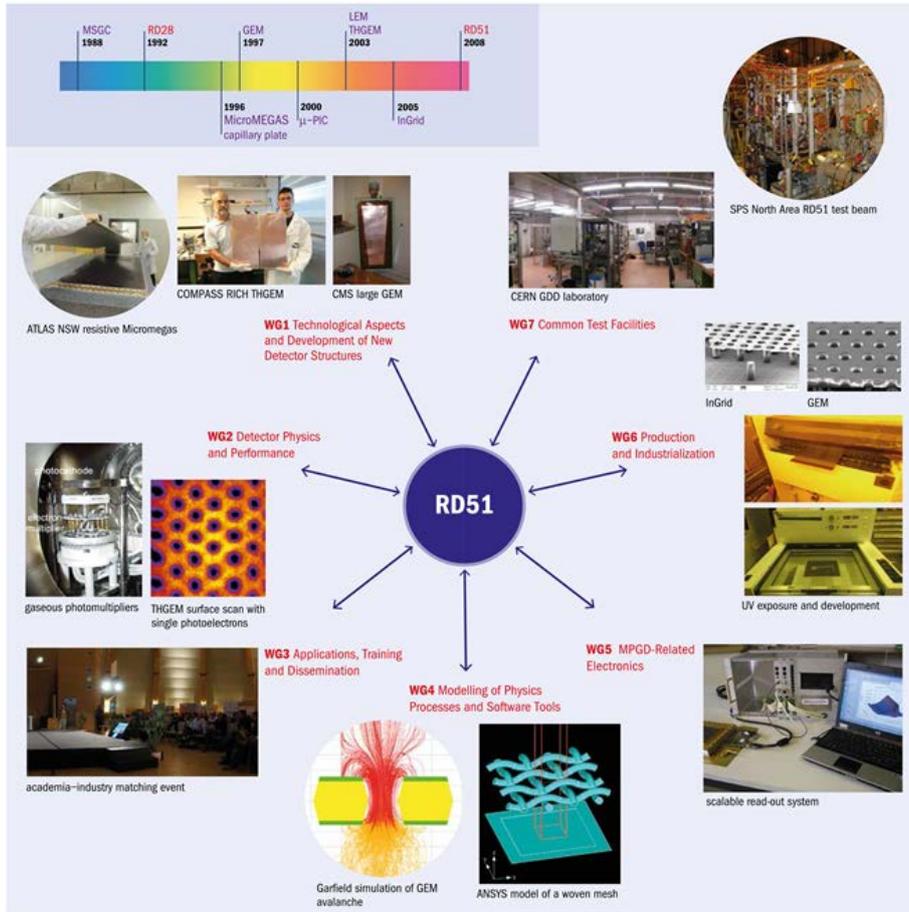

**Figure 4.** The seven working groups of RD51 Collaboration "Development of Micro-Pattern Gaseous Detector Technologies", consisting of more than 80 institutes world-wide, with illustrations of a few examples of the different kinds of work involved. [39,40]

The Worldwide LHC Computing Grid (WLCG) was established in 2001 to address a major computing challenge for the particle physics community; today it includes resources from more than 170 sites in 42 countries with > 2 million/jobs per day. The WLCG was developed for the LHC experiments, but Belle II, LBNF/DUNE and the linear collider detector studies use the same infrastructure. The challenges to face were (and still are): to keep operating the system reliably enough with significantly less effort, to evolve towards new, more open and flexible computing models, and to provide the software and the resources needed by the LHC experiments. In terms of data processing, the rate of advances in hardware performance has slowed in recent years, encouraging the community to adapt and to take advantage of developments such as graphics processing unit (GPU), high-performance computing (HPC) and commercial Cloud services [44]. Cloud computing is a technology largely viewed as the next big step in the development and deployment of an increasing number of distributed applications. Although commercial Clouds are not yet economically viable for large scale data intensive HEP computing, the particle physics community should seize the opportunity to be involved in the planning stage of future multidisciplinary research infrastructures. Efforts to employ HPC centers, based on GPU architectures, and big-data techniques from industry are ongoing to achieve record breaking results (towards exascale). The LHC computing infrastructure so far has been largely based exclusively on X86 architecture using CPUs, while GPU are gaining a lot of popularity as co-processors due to the success of Machine Learning and "Artificial Intelligence". A broad overview of machine learning applications, research and development directions in particle physics can be found in the High-Energy Physics Machine Learning Community White Paper [45]. In order to address future challenges, all LHC experiments are planning to use GPUs at some level in their future upgrades: ALICE will employ a GPU based Online/Offline system; CMS is porting part of their trigger software to run on GPU processors; LHCb is exploring GPUs for their online data reduction [46]; ATLAS is developing algorithms to run on GPUs [12]. The HEP software foundation released a community white paper in 2018 setting out the radical changes in computing and software – not just for processing but also for data storage and management – required to ensure the success of the LHC and other particle physics experiments into the 2020s [47].

## 3. Advanced Concepts in Vertex and Tracking Detectors

Key elements for most experiments in high energy and nuclear physics are vertex and tracking detectors. They are used to determine the charge, momentum, and energy of traversing particles and to allow quark-flavor and tau-lepton identification through the reconstruction of secondary vertices. Gaseous and semiconductor detectors are the two main types of tracking detectors; other, more exotic ones are fiber-based (e.g. LHCb tracker upgrade) or transition radiation tracking (e.g. ATLAS TRT) devices [48]. While gaseous detectors offer sizeable low-mass volumes and many measurement points for an excellent pattern recognition and ultimate dE/dx measurement (e.g. cluster counting technique can be exploited instead of the charge-analog information, as discussed in Section 3.2), the silicon-based approach offers the most accurate single point resolution. The challenge of the most intense tracking regions at HL-LHC will be addressed by hybrid pixel detectors, and the breakthrough technology is expected to come from CMOS sensor developments, which are currently being pushed to unprecedented levels of radiation hardness and rate capabilities. For future applications in particle physics, alternative technologies, by employing beyond state-of-the-art interconnection technologies, such as 3D vertical integration, through-silicon-vias (TSV), or micro bump-bonding, which, while retaining the advantages of separate and optimized fabrication processes for sensor and electronics, would allow fine pitch interconnect of multiple chips.

### 3.1 Solid State (Silicon) Detectors

Silicon detectors have transformed the field of particle physics over the previous four decades. Their rise, originally started in the CERN's NA11 and NA32 fixed target experiments to study lifetime of charmed mesons, has followed a version of "Moore's law" expansion (approximate doubling in the number of transistor per mm$^2$ every ~two years over previous decades) both in terms of surface covered and number of readout channels (see Figure 5 [49]). Since then, silicon detectors have been engineered in increasingly large and complex systems (e.g. silicon-strips at ALEPH, DELPHI, OPAL, L3, BABAR, Belle, CDF, DO, hybrid-pixels at WA97 and DELPHI, charged-coupled device (CCD) at SLD, and NA62 tracker with per-pixel timing of ~150 ps). The readout electronics is typically fabricated in a commercial CMOS process, and directly connected to the sensors via wire- or bump-bonding. The geometry of the sensors vary significantly for different applications, as shown in Figure 6, but silicon detectors are always made of three functional blocks: the sensing diode, the analogue amplifier and the digital circuit. The main future directions are lower cost production of radiation-hard hybrid pixels and planar-diode sensors and development of monolithic devices incorporating complex readout architectures in CMOS foundries. Fast picosecond-time sensors based on Low-Gain Avalanche Detectors (LGAD) (as discussed in Section 4) and 3D-devices will be also exploited.

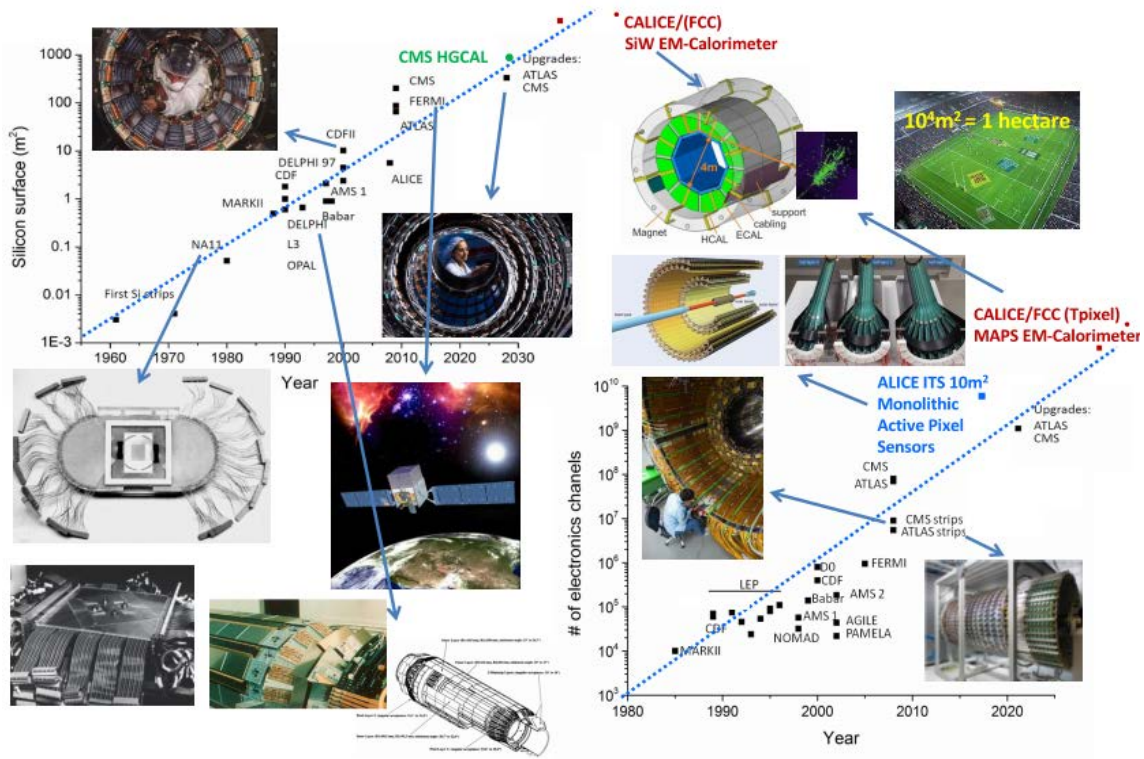

**Figure 5.** Progress in both the area of silicon and number of electronics channels in detector arrays with construction date (by P. Allport [49]).

The LHC vertex/tracking silicon detectors represent the state-of-the-art technology and have reached an excellent level of maturity at an industrial level together with high radiation tolerance up to ~ $10^{16}$ $n_{eq}/cm^2$, particle rates up to GHz/cm$^2$ and pixel sizes down to ~ 50x50 µm$^2$. Today, all four LHC experiments are engaged in major upgrades [50]. During LS2, LHCb plans to upgrade the VELO detector and ALICE will upgrade the Inner Tracking System (ITS) using Monolithic Active Pixel Sensors (MAPS), as shown in Figure 7. Further down the road, in LS3, CMS and ATLAS plan major upgrades of their trackers, including a combination of pixel and strips sensors for the inner and outer layers. This will enlarge a total area of silicon sensors at HL-LHC to ~1000 m², including CMS HGCAL calorimeter (discussed in Section 6), ALICE FoCAL calorimeter, and ATLAS HGTD and CMS ETL timing detectors (discussed in Section 4), representing a significant challenge for procurement. In general, pixel detector systems will enlarge dramatically in surface (ATLAS by factor of ~15) and in channel count (ALICE will reach 12.5 billion pixels with CMOS MAPS). The LHCb VELO will be upgraded to all-pixel system (cell size will decrease by a factor of ~ 1000), while the Expression of Interest envisages a "5D"-tracking system with a modest per-hit timing information (~50-200 ps) for every single track point in the vertex detector to improve association of tracks to the correct collision vertex [51]. Such a device would be of interest across LHC experiments.

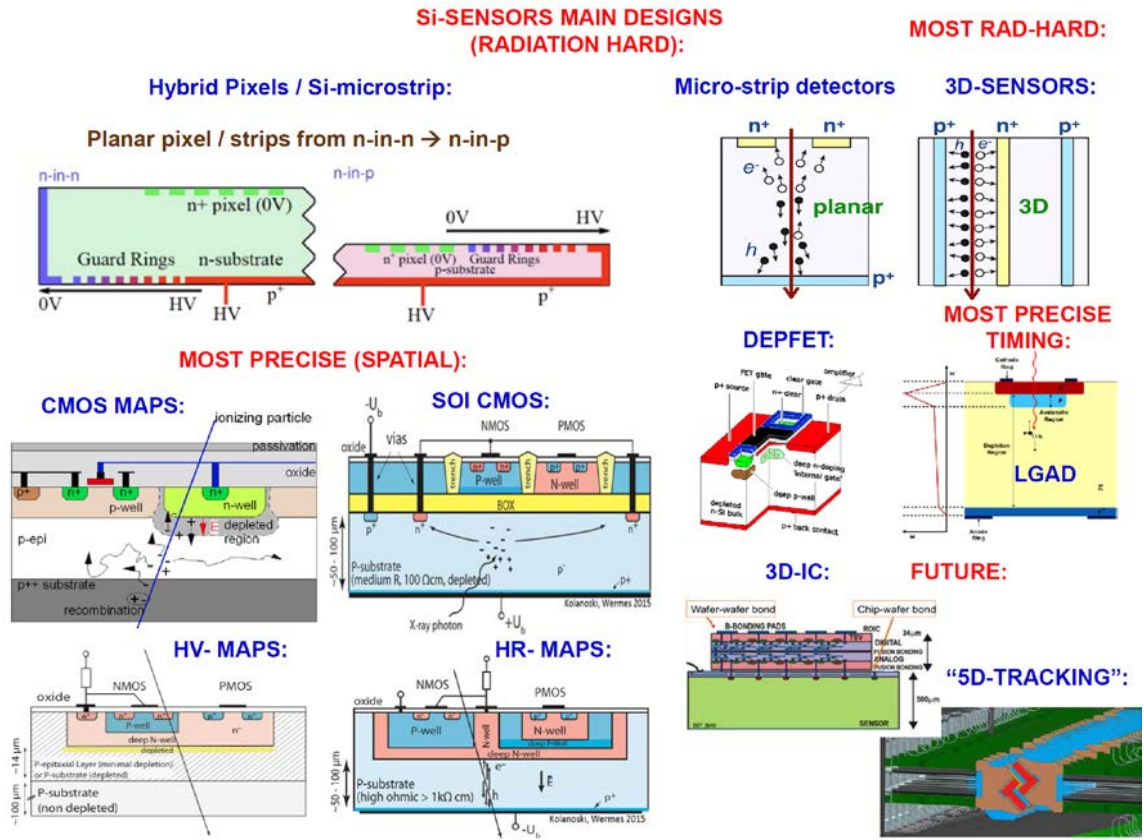

**Figure 6.** The landscape of modern tracking detector technologies based on solid-state (silicon) devices.

All present LHC experiments (ATLAS, CMS, ALICE and LHCb for RICHs with pixel-HPDs) employ the hybrid pixel technique for large systems O(~m²), which represents an admirable achievement in complex engineering and integration. The baseline solution for LHC sensors has n+-implants on n-type substrate material (n-in-n) in the ATLAS, CMS and LHCb vertex locator (VELO), kept at sub-zero temperature. It allows operation in partial depletion after type inversion when the remaining depletion zone grows from the segmented front of the sensor (in contrast to p-in-n). Currently, the trend in hybrid pixel modules for HL-LHC has been tailored towards thin sensors (100-150 μm) using n+ pixel implants in high-quality p-substrate material (>2kΩcm) having a faster collection time. With (n-in-p) type sensors cost-efficient single sided processing suffices for fabrication, compared to double-sided for (n+-in-n), while maintaining electron collection (n-type pixels). Thin 6- or even 8-inch sensor wafer production is enabled by techniques employing SOI or Si-Si handling wafers, or by thinning (e.g. by cavity etching); the current sensor thickness limit is around 50 µm [52]. The bump bonding may be still a limiting approach and investment is necessary on emerging interconnect technologies (small pitch bump bonding, TSV) enabling assembly at extremely small pixel pitches O(~10 µm).

| Exp. / Timescale | Application Domain | Tech. | Detector size / Module size / Channel count | Radiation Environment | Special Req. / Remarks |
|---|---|---|---|---|---|
| ATLAS ITK Upgrade CERN LS3 | Hadron Collider (Vertex / Tracking) | Si hybrid pixels (n-in-p), 3D innermost, Si-Strips | Total area: pixel – 12.7 m$^2$ ; strips - 165 m$^2$ Single unit: pixel- 50x50 (25x100) µm$^2$ strip len./pitch: ~24 – 80 mm / ~70 µm Channels count : pixels – 5 G ; strips – 60 M | Fluences up to 2 x 10$^{16}$ n$_{eq}$/cm$^2$ | Option for outermost pixel layer: MAPS RD53 ASIC 65 nm CMOS |
| CMS Tracker Upgrade CERN LS3 | Hadron Collider (Vertex / Tracking) | Si hybrid pixels (n-in-p), 3D innermost, Si-Strips | Total area: pixel - 4.9 m$^2$ ; strips - 200 m$^2$ Single unit: pixel- 25x100 (50x50) µm$^2$ strip len./pitch: 50-24-1.5 mm /~100 µm Channels count : pixels – 3 G ; strips – 175 M | Fluences up to 2.3 x 10$^{16}$ n$_{eq}$/cm$^2$ | Special p$_T$-modules in outer strip layers RD53 ASIC 65 nm CMOS |
| ALICE ITS Upgrade CERN LS2 | Heavy Ion Physics (Tracking) | CMOS MAPS, 7 barrel layers | Total area: 10 m$^2$ ; Single unit: pixel size 30x30 µm$^2$ Channels count : 12.5 G | Fluences up to 1.7 x 10$^{13}$ n$_{eq}$/cm$^2$ | 0.3% X$_0$ per layer (inner barrel) ASIC: 180 nm TowerJazz |
| LHCb VELO Upgrade CERN LS2 | Hadron Collider (B Physics) | Si hybrid pixels (n-in-p) | Total area: 0.12 m$^2$ ; Single unit: pixel size 55x55 µm$^2$ Channels count : 41 M | Fluences up to 8 x 10$^{15}$ n$_{eq}$/cm$^2$ | 130 nm CMOS, 40 MHz VELOPIX readout, rates up to 20 Gb/s |
| LHCb Upstream Tracker Upg. CERN LS2 | Hadron Collider (B Physics) | Si strips (n-in-p & p-in-n) | Total area: 9 m$^2$ ; Single unit: strip length/pitch: 50 -100 mm / 100 – 200 µm$^2$ Channels count : ~ 500k | Fluences up to 5 x 10$^{14}$ n$_{eq}$/cm$^2$ | |
| BELLE II PXD / SVD | e+e- Collider (B Physics) | DEPFET / Si-strips (p-in-n) | Total area: 0;03 m$^2$ / 1.2 m$^2$ ; PXD unit: pixel size ~50x50 µm$^2$ SVD unit: strip- 120 mm / 50-240 µm$^2$ Channels count : 7.7 M / 245 k | Fluences up to 10$^{13}$ n$_{eq}$/cm$^2$ | 0.15 X$_0$ per layer |

**Figure 7.** Summary of the vertex/tracking silicon detector systems at HL-LHC and Belle II. Updated plot from C. Joram [53].

3D-silicon detectors (3D) are based on the implementation of vertical column electrodes penetrating inside the detector bulk. In such configuration, the charge drifts parallel to the sensor surface, as opposed to the vertical drift in planar ones, and the distance depends on the column separation (typically ~ 50 µm). Due to their design, the main detector figures of merit are low depletion voltage (i.e., low power dissipation), charge drift decoupled from particle track (allowing short collection time), low drift path (so, reduced trapping) and reduced charge sharing. Their development, nowadays carried out by two main producers (FBK and CNM), has been mainly focused on the innermost ATLAS-IBL pixel layers plus several joint MPW production runs with CMS and LHCb. Despite their quite complicate fabrication process, that can be both double- and single-sided, 3D detectors are the most radiation-hard technology to-date demonstrating good performance even up to a fluence of ~ 3 x 10$^{16}$ n$_{eq}$/cm$^2$ and time resolution ~30 ps at V$_{bias}$ > 100V and T = -20C [54]. New developments are being pursued to address the challenges of radiation hardness, small pitch with narrower columns, thinner devices, better breakdown behaviour, improved on-wafer selection and lower costs.

The Si-strip sensors in the upgraded ATLAS/CMS trackers will consist of n-type strips in p-type substrates (n-in-p) and replace the previously used (p-in-n) technology. This is driven by radiation hardness consideration, that ionizing and displacement damage have a less detrimental impact on detector performance for n-in-p devices than for p-in-n devices. In particular, the future

ATLAS ITK strip system will use n+-in-p float zone (FZ) silicon of 310 μm thickness and comprise of ~ 18,000 modules with 60 million channels, covering an area of 165 m$^2$ [55].

Another novelty is the use of CMOS Pixel Sensors (CPS) in a hadron collider, which have been developed since two decades for an ILC vertex detector [56]. Initially deployed in test-beam infrastructure (EUDET telescope) and later adopted in the STAR HFT (~20 x 20 μm² pixels) and ALICE ITS (< 30 x 30 μm² pixels), they are considered now for all future e+e- "Higgs Factories" and also for high-rate applications like the ATLAS pixel detector upgrade [57]. The ALPIDE architecture for ALICE ITS allows for full 180 nm CMOS in-pixel circuitry by shielding the n-well containing PMOS transistors with a deep p-well implant from the epitaxial layer preventing them from collecting signal charge and removing the limitation on time resolution set by the read-out speed. This approach leads to a power consumption of less than 40 mW/cm$^2$, a spatial resolution of around 5 μm, and a radiation tolerance of ~$10^{13}$ $n_{eq}$/cm$^2$, fulfilling or exceeding the ALICE requirements [58]. As an extension of ALPIDE design, large-scale MIMOSIS prototype is being developed for the FAIR CBM experiment; the latter could serve as a possible baseline prototype for ILC vertex detector. The R&D pursued so far did not show that a spatial resolution of ≲ 3 μm and a single bunch tagging at ILC are actually achievable simultaneously within the same, 50 μm thin, sensor with the presently available CMOS technologies and a power consumption compatible with air-cooling. However, ~4 μm spatial resolution may be within reach with CPS sensors featuring a continuous read-out time of ≲ 1 μs at an affordable power consumption [32]. An alternative CPS approach is the 180 nm Chronopixel design with a few hundred nanosecond time-stamping capability [59]. A recent trend of exploiting the high-voltage/high-resistivity (HV/HR) add-ons of CMOS technologies to increase depletion of the sensing volume, also known as Depleted MAPS (DMAPS), has enabled fast charge collection by drift in the sensors [60]. For near future applications, large prototype chips have become available in CMOS foundries (LFoundry 150 nm, AMS 180 nm, and the modified TowerJazz 180 nm process) [61]. Current CMOS pixel technology can qualify for the HL-LHC requirements, e.g. the outermost pixel layer of ATLAS ITK, in terms of radiation hardness up to levels of ~ $10^{15}$ $n_{eq}$/cm$^2$ [62], but require more design work to meet all specifications.

The radiation tolerance is a major challenge for vertex and tracking systems in the current and future hadron colliders. The main degradation mechanism is a displacement damage when a particle impact knocks an atom out of its lattice position, followed by a variety of interactions of the created vacancy and interstitial, leading to larger leakage currents, enhanced trapping of signal charge, and changes in effective doping, which can also strongly evolve during annealing. The new extreme radiation environment (anticipated fluence in the ATLAS/CMS experiments at HL-LHC are ~ $2\times10^{16}$ $n_{eq}$/cm$^2$ ($2\times10^{15}$ $n_{eq}$/cm$^2$) for innermost pixel layers (micro-strip sensors) instrumenting the tracker and going up to ~ $8\times10^{17}$ $n_{eq}$/cm$^2$ for FCC-hh) called for another leap in the field of silicon sensors and microelectronics technologies. Radiation hard device engineering includes silicon with oxygen-diffused FZ content supplied in the growth process, which is proved to have superior radiation hardness against charged particles, but not neutrons [63,64]. In line with this, a better understanding of radiation damage effects became a research branch on its own, in the context of RD50 for sensors and microelectronics [28], including activities within RD53 [30]. It is clear that silicon will remain the choice for the years to come, but whether these detectors can stand the FCC-hh fluence is yet to be seen.

Vertex detectors at a future electron-positron "Higgs Factory" (ILC, CLIC, FCC-ee or CEPC) will operate in an environment with high (continuous or bunched) beam currents, a minimum distance from beam axis of about 15 mm, a requirement of ≤5 μm single point resolution, high granularity ($\leq 30 \times 30$ μm$^2$), power dissipation ($\leq 50$ mW/cm$^2$), low mass (~ 0.1% of $X_0$, or 100 μm Si-equivalent per layer) and a moderate radiation tolerance ~$10^{11}$ n$_{eq}$/cm$^2$ per year. The clear challenge is unprecedented spatial resolution, to be achieved with ultra-small pixels, and thus extremely low material budget. Very thin detector assemblies are mandatory, while providing high stiffness and stability. Heavy ion experiments, B-factories, nuclear physics and rare muon decay experiments also operate in a multiple-scattering dominated regime. The distinct ILC and CLIC beam structures with its low duty cycle allows for triggerless readout between the trains and operation with the power off during the inter-train periods (power pulsing) to reduce on-detector cooling needs. Both ILC and SiD concepts have in common that they are based on monolithic (CMOS, Chronopix, SOI), DEPFET or FPCCD devices, since LHC-like hybrid structures impose too much material. A new vertex detector comprising a two-layer pixel detector based on DEPFET concept has been developed, including solutions for mechanical support, cooling and services and partially installed in the Belle II experiment [65]. Extending this concept to the ILC requires re-optimizing the trade-off between pixel dimensions, sensor thickness and readout time. The recent developments of Silicon-On-Insulator (SOI) technology is impressive; it allows production of thin monolithic sensors consisting of a CMOS readout separated from a fully depleted high-resistivity sensor layer through an insulator oxide layer [66]. Very interesting is the use of additional implants to shape the lateral field in the sensor for better timing performance. The main assets of SoI device for ILC rely on the high-density in-pixel circuitry, particularly with the double-tier option recently addressed. The main specifications of the CLIC detector include a small pixel size (25x25 μm$^2$), simultaneous energy and time measurement (~O(10 ns) accuracy) and power consumption of ~ 50 mW/cm$^2$, while keeping material budget low; in this way, air cooling solution could be envisaged. Recent studies have been performed with a small-feature size integrated hybrid approaches (50 μm Si-sensors with Timepix or CLICpix2 ASIC), HV-CMOS/HR-CMOS process technologies or SOI [35,67]. Both high-density and low power consumption will benefit from an upgrade to a 65 nm technology. Finally, development of a new type of hybrid vertex detector for CLIC, called Enhanced Lateral Drift (ELAD) sensor, where position resolution is improved by a dedicated charge sharing mechanism, was reported in [68].

A future silicon pixel tracker, which excludes from the detector acceptance all services and mechanical support structures would represent an ideal design. Thicknesses of the sensors down to 50μm can be achieved, beyond which mechanical supports, readout bus-tapes and cooling services typically dominate. To minimize material budget, new technologies, like stitching, will allow developing a new generation of large-size CMOS MAPS with an area up to the full wafer size. In this technology, the reticles which fit into the field of view of the lithographic equipment are placed on the wafer with high precision, achieving a tiny but well defined overlap. In addition to large-size sensors, it may also be useful to bend thin (50 μm) sensors to make cylindrical assemblies. This would allow the construction of vertex detectors with unprecedented low values of material thickness (<0.1% $X_0$), confining the electrical interconnection to the sensors edges, uniform coverage and very high spatial resolution (~ 3 μm). Stitching is available in the TJsc 180 nm process used for ALPIDE and MIMOSIS. It is also accessible in a forthcoming 65 nm imaging process investigated for the purpose of the ALICE upgrade. Good relations with the IC foundries are essential for the future development of this technique [56,69,70]. The use of CMOS and stitching technologies for silicon pixel devices will open new opportunities in vertex detectors.

Low mass mechanics for sensors support and thermal management represents a major challenge for future tracking detectors. Common ATLAS and CMS strategies to reduce material budget are: implement fewer layers, use of DC-DC converters in the outer strip tracker, serial power in the inner pixel tracker, ultra-light structural materials, $CO_2$ cooling, etc [68]. The current state-of-the-art for inner tracking detectors at the HL-LHC is two-phase evaporative cooling with $CO_2$ fluid and a target "on-detector" evaporation temperature is -40°C. Micro-channels embedded in silicon or polyimide substrate for vertex detector for both single/two-phase cooling fluids represent a powerful solution [71]. In particular, $CO_2$ micro-channel cooling will be deployed as a real novelty in the LHCb VELO upgrade. At the extreme of the FCC-hh, the harsh radiation environment will impose severe constraints on the mechanical design and materials suffering from loss of mechanical, electrical and optical properties, while at the same time driving the cooling design. The silicon microfabrication processes or carbon composite microvascular approaches will need to be fully exploited and new additive manufacturing technologies (3D printing) and materials will be investigated [26]. At Linear Colliders, the cooling system will depend on the technology choice. The baseline consists in air-cooling which is expected to be able to extract the total power dissipation of the vertex detector (few tens of watts). More specific developments are also being pursued (e.g. micro-channel cooling for DEPFET, two-phase $CO_2$ cooling system for Fine Pixel CCDs, etc.).

**3.2 Gaseous Detectors**

More than fifty years ago, in 1968, the instrumentation domain was revolutionized by G. Charpak's invention of the Multi-Wire Proportional Chamber (MWPC), earning him the 1992 Nobel Prize in Physics. Crucially, the emerging integrated-circuit technology could deliver at that time amplifiers small enough in size and cost to equip many thousands of proportional wires. This invention marked the transition from optically read out detectors, such as bubble and cloud chambers, to the electronic era. Confronted by the increasing demands of particle physics experiments, numerous advanced concepts with novel geometries and exploiting various gas properties have been developed, such as Drift Chamber (DC), Time Projection Chamber (TPC), Multi-step Avalanche Chamber, Ring-Imaging Cherenkov Counter (RICH), Resistive Plate Chamber (RPC), and many others [72]. In the original design, the MWPC/DC consists of a set of parallel, evenly spaced, anode wires stretched between two cathode planes; the position of a track can be estimated by exploiting the arrival time of electrons at the anodes if the time of interaction is known. Wire-based drift chambers have become well established in the fertile field of gaseous detectors during the past 40 years [73] and are acknowledged as highly performing tracking devices successfully exploited in many recent experiments (e.g. Belle II CDC [74]). An ultra-light weight drift chamber for high precision momentum reconstruction and particle identification is under investigation for the future FCC-ee and CePC colliders. Its design was inspired by the DAFNE's KLOE large wire chamber as well as by a more recent version of it developed for the MEG2 experiment [75]. Conventionally, drift chambers have been operated with hydrocarbon-based mixtures, which are not trustable for the long-term, high-rate operation [76]. Hence, a dedicated study might be necessary to find an alternative hydrocarbon-free mixture adapted to the desired DC performance at future colliders. Despite various improvements, position-sensitive detectors based on wire structures are limited in terms of maximum rate capability and detector granularity, due to the space-charge effects, for the LHC-like high-rate environments.

The "ultimate" drift chamber is the TPC concept, which combines a measurement of drift time and charge induction on the endplates to achieve excellent pattern recognition for high multiplicity environments and moderate rates; examples are future lepton colliders or heavy-ion experiments. This technology provides 3D precision tracking; the gaseous detector volume gives an extremely low material budget; and the high density of space points enables particle identification (PID) through ionization loss (dE/dx) measurement. A further improvement comes from a more accurate description of the ionisation energy loss by the so-called Bichsel functions. An important major innovation in the design of the future TPCs is related to the replacement of MWPCs with Micro-Pattern Gas Detectors (MPGDs) for the endplate readout detectors.

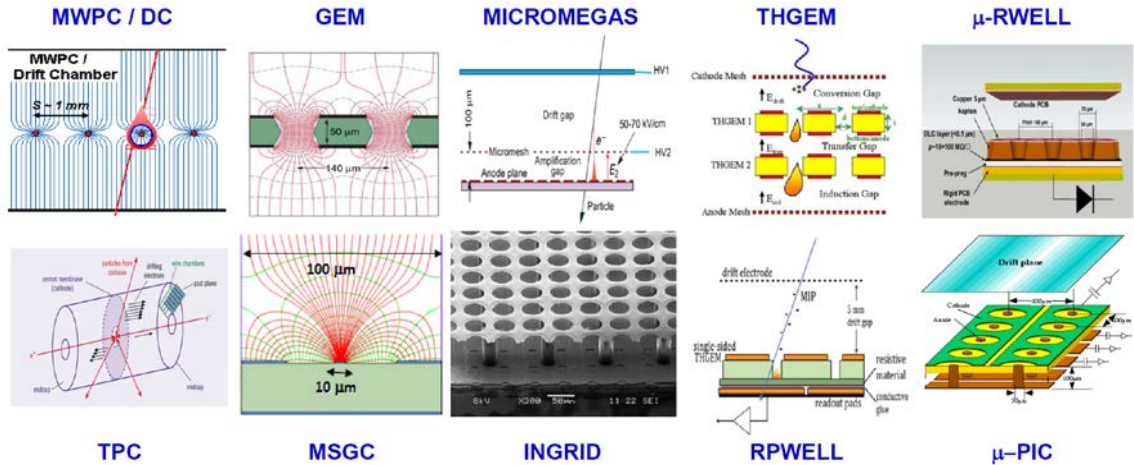

**Figure 8.** Gaseous Detectors Family: from Multi-Wire Proportional and Drift chambers (MWPC/DC) to the Time Projection Chamber (TPC) and ultimately to the modern Micro-Pattern Gaseous Detector (MPGD) technologies.

Similar to silicon technology and the IT domain, industrial advances in photolithography, microelectronics and printed circuit boards (PCB) has favored the invention of novel micro-structured gas-amplification devices. Nowadays, a broad family of MPGD technologies are being developed/optimized, such as: Gas Electron Multiplier (GEM), Micro-Mesh Gaseous Structure (MM), THick GEMs (THGEM), also referred to in the literature as Large Electron Multipliers (LEM), Resistive Plate WELL (RPWELL), GEM-derived architecture (μ-RWELL), Micro-Pixel Gas Chamber (μ-PIC), and an integrated readout of gaseous detectors using solid-state pixel chips (InGrid), as shown in Figure 8. By using pitch size of a few hundred microns, an order of magnitude improvement in granularity over wire chambers, these devices exhibit intrinsic high rate capability (up to $10^6$ Hz/mm$^2$), excellent spatial resolution (down to 30 μm), and single photo-electron time resolution in the ns-range [77]. Nevertheless, the integration of MPGDs in large experiments was not rapid, despite of the first large-scale application within the COMPASS experiment at CERN SPS in the 2000's [78]. The recent choice of MPGDs for relevant upgrades of CERN experiments indicates the degree of maturity of given detector technologies, the level of dissemination within the particle physics community and their reliability. A big step in the direction of the industrial and cost-effective manufacting of MPGDs was the development of the new fabrication technologies – resistive MM [79], to suppress destructive sparks in hadron environments, single-mask and self-stretching GEM techniques [80], which enable production of large-size foils and significantly reduce detector assembly time. Scaling up MPGD detectors, while preserving the typical properties of small prototypes, allowed their use in the LHC experiments after LS2 (Micromegas [81] in combination with Thin Gap chambers [82] in the ATLAS New Small Wheel, GEMs in the CMS Muon System [83] and in the ALICE TPC). In addition, wire-chamber based photon detectors of COMPASS RICH-1 have been replaced by hybrid MPGDs - staggered THGEM and a MM multiplication stage with CsI photocathodes - to accomplish the delicate mission of single photon detection (see Section 5). Figure 9 summarizes the use of MPGDs in the recent CERN experiments and HL-LHC upgrades. In addition, MPGDs are currently being implemented in a variety of nuclear physics experiments and also foreseen for future facilities (e.g., ILC, CePC [84,85], FCC, EIC, FAIR and NICA [86]); as an example, muon detectors in FCC experiments will cover an active areas larger than 1000 m$^2$. Large R&D efforts on TPCs are ongoing for the ALICE and MPD experiments [87,88], as well as for the future ILC and CePC colliders. In particular, the ILC beam bunch structure allows the implementation of the gating scheme for the ILD detector, based on large-aperture GEMs with honeycomb-shaped holes [89], to minimize deteriorating influence of ion backflow on spatial resolution. Three readout options are being developed for the TPC at ILC: GEM [90], MM [91] and InGrid [92]. In contrary, under Z-pole operation mode at the CePC upgrade (L~$10^{36}$ cm$^{-2}$s$^{-1}$), there might be difficult to implement "open/close" gating mode scheme due to the lack of time. In this case, MPGD associated to the silicon pixel technology (e.g. "InGrid" concept) is a promising option for the CePC and ILC. In fact, this innovative technique provides the high granularity needed to resolve individual clusters which are separated by an average distance of a few hundred microns. The goal is to determine dE/dx by cluster counting technique, rather than by measuring the charge, with an ultimate precision of better than 3%. Moreover, this approach provides an unprecedented potential for pattern recognition in dense environments leading to superior double hit/track resolution and the possibility to discriminate against δ-rays.

| Experiment / Timescale | Application Domain | MPGD Technology | Total detector size / Single module size | Operation Characteristics / Performance | Special Requirements / Remarks |
|---|---|---|---|---|---|
| COMPASS TRACKING > 2002 | Fixed Target Experiment (Tracking) | 3-GEM Micromegas w/ GEM preampl. | Total area: 2.6 m$^2$ Single unit detect: 0.31x0.31 m$^2$ Total area: ~ 2 m$^2$ Single unit detect: 0.4x0.4 m$^2$ | Max.rate: ~100kHz/mm$^2$ Spatial res.: ~70-100µm (strip), ~120µm (pixel) Time res.: ~ 8 ns Rad. Hard.: 2500 mC/cm$^2$ | Required beam tracking (pixelized central / beam area) |
| TOTEM TRACKING: > 2009 | Hadron Collider / Forward Physics (5.3≤|η| ≤ 6.5) | 3-GEM (semicircular shape) | Total area: ~ 4 m$^2$ Single unit detect: up to 0.03m$^2$ | Max.rate: 20 kHz/cm$^2$ Spatial res.: ~120µm Time res.: ~ 12 ns Rad. Hard.: ~ mC/cm$^2$ | Operation in pp, pA and AA collisions. |
| LHCb MUON DETECTOR > 2010 | Hadron Collider / B-physics (triggering) | 3-GEM | Total area: ~ 0.6 m$^2$ Single unit detect: 20-24 cm$^2$ | Max.rate: 500 kHz/cm$^2$ Spatial res.: ~ cm Time res.: ~ 3 ns Rad. Hard.: ~ C/cm$^2$ | Redundant triggering |
| COMPASS RICH UPGRADE > 2016 | Fixed Target Experiment (RICH - detection of single VUV photons) | Hybrid (THGEM + CsI and MM) | Total area: ~ 1.4 m$^2$ Single unit detect: ~ 0.6 x 0.6 m$^2$ | Max.rate: 100 Hz/cm$^2$ Spatial res.: << 2.5 mm Time res.: ~ 10 ns | Production of large area THGEM of sufficient quality |
| ATLAS MUON UPGRADE CERN LS2 | Hadron Collider (Tracking/Triggering) | Resistive Micromegas | Total area: 1200 m$^2$ Single unit detect: (2.2x1.4m$^2$) ~ 2-3 m$^2$ | Max. rate: 15 kHz/cm$^2$ Spatial res.: <100µm Time res.: ~ 10 ns Rad. Hard.: ~ 0.5C/cm$^2$ | Redundant tracking and triggering; Challenging constr. in mechanical precision |
| CMS MUON UPGRADE CERN LS2 | Hadron Collider (Tracking/Triggering) | 3-GEM | Total area: ~ 143 m$^2$ Single unit detect: 0.3-0.4m$^2$ | Max. rate: 10 kHz/cm$^2$ Spatial res.: ~100µm Time res.: ~ 5-7 ns Rad. Hard.: ~ 0.5 C/cm$^2$ | Redundant tracking and triggering |
| ALICE TPC UPGRADE CERN LS2 | Heavy-Ion Physics (Tracking + dE/dx) | 4-GEM / TPC | Total area: ~ 32 m$^2$ Single unit detect: up to 0.3m$^2$ | Max.rate: 100 kHz/cm$^2$ Spatial res.: ~300µm Time res.: ~ 100 ns dE/dx: 11 % Rad. Hard.: 50 mC/cm$^2$ | - 50 kHz Pb-Pb rate; - Continues TPC readout - Low IBF and good energy resolution |

**Figure 9.** Summary of the Micro-Pattern Gaseous Detector Technologies employed/proposed for CERN Experiments.

The consolidation of the better-established technologies has been accompanied with flourishing of novel ones, often specific to well-defined applications. Modern technologies have been also derived from MM and GEM concepts, hybrid approaches combining different elements in a single device, gaseous with non-gaseous detectors, as it is the case for optical read-out [93,94]. MPGD hybridization, a strategy aiming to strengthen the detector performance, remains a valid asset for addressing future experimental challenges such as high granularity and precision timing. As a result, all flavours of MPGDs are in high demand for future applications in particle and nuclear physics, including cryogenic LAr/LXe detectors for neutrino physics and dark matter searches. Hence, industrial manufacturing became mandatory and remains a central issue to be solved. Better understanding of the physics processes, originating from experiment-validated model simulations, paved the road towards novel detection concepts. A clear direction for future developments is that of resistive materials and related detector architectures [95,96]. Their usage improves detector stability, making possible a higher gain in a single multiplication layer, a remarkable advantage for assembly, mass production and cost. Diamond-like carbon (DLC) resistive layers are the key ingredients for increasing the rate capability of MPGDs [97,98]. Nowadays, many intensive R&D activities and their diversified applications are pursued within the world-wide CERN-RD51 collaboration [99].

Future developments call for novel materials as well as for new fabrication techniques. Contributions to the detector concepts are required for several domains: resistive materials, solid-state photon and neutron converters, innovative nanotechnology components. Material studies can contribute to requirements related to low out-gassing, radiation hardness, radio-purity, converter robustness and eco-friendly gases. The development of the next generation of MPGDs can largely profit of emerging technologies as those related to MicroElectroMechanical Systems (MEMS), sputtering, novel photoconverters, 3-D printing of amplifying structures and cooling circuits, etc. This activity will enlarge community expertise in the domains of nanotechnologies and material science, fields that are not exhaustively covered today and which might bring large potential for future applications. This is just a partial list of fascinating R&D lines that MPGDs will see in the years to come. Gaseous detectors are flexible and wide-spread devices and will remain a key technology in particle physics experiments.

**3.3 Scintillating Fiber Trackers**

Scintillating fiber (SciFi) tracking combines the speed and efficiency of a scintillation detector with the flexibility and hermeticity afforded by fiber technology. Large fiducial volumes can be instrumented with the particle-sensitive fiber arrays conformed to custom-designed surfaces; novel architectures are also possible. The SciFi technology was for the first time applied at large scale in the UA2 detector upgrade and later adopted in the CERN CHORUS experiment. The Central Fibre Tracker (CFT) in D0 experiment at Fermilab, coupled to waveguide fibers and readout with a solid-state visible light photon counters (VLPC), marks another milestone [100]. The small size and their insensitivity to magnetic fields makes silicon photomultipliers (SiPM) an attractive option for a scintillator-based tracking system [101]. This technology has been adopted for the upgrade of the LHCb experiment: a large scintillating fiber tracker readout by SiPM array with a total active area of 340 $m^2$ [102]. Another example is the novel SuperFGD concept for the T2K upgrade (see Section 9).

However, fibre technology itself has not seen much innovation in the past decades and their modest radiation hardness is a limitation. The cumbersome and very labour-intense production adventure of fibre detectors may also be seen as a bottleneck. While silicon and more recent MPGDs have been produced by highly precise and automated microfabrication technologies, including photo-lithographic patterning, there is no obvious alternative to the traditional drawing, winding, casting and machining of fibres. It is possible that the evolution of the 3D-printing technique will pave new road for this technology [103].

## 4. Advanced Concept in Picosecond Timing Detectors

Precision timing detectors in the (sub)-nanosecond range have been extensively used in high-energy physics experiments either for triggering purposes, to attribute the correct timing to the collision event (bunch crossing information), or to identify particle mass, by combining the time-of-flight and momentum measurements. Driven by the general push towards higher luminosities not only at HL-LHC, but also at Belle II experiment, and the next generation of colliders, timing information could bring a major benefit by allowing to implement 4D reconstruction of the primary collision vertex - joint pattern recognition, based on timing and tracking detectors, enables imaging of the event topology. In particular, the required accuracy in the ATLAS and CMS at HL-LHC will be dictated by the pile-up induced backgrounds and the nominal interaction distribution within a single bunch crossing ($\sigma_z \sim$ 5 cm, $\sigma_t \sim$ 170 ps rms). A typical time resolution of a few tens of picosecond will add a new handle in associating photons with individual vertices at the LHC and rejection of spurious energy deposits in the calorimeter that are not consistent with the primary vertex time [104,105]. Ultra-fast timing in calorimetry can be also used to resolve the development of hadron showers, by separating their electromagnetic and hadronic components, and therefore simplifying the implementation of the particle flow algorithm. In general, space–time tracking could be used in kaon rare decay experiments, many physics analysis at LHC – Higgs, BSM searches for long-lived particles, by measuring precisely the time-of-flight between their production and decay, and/or in assigning beauty and charm hadrons to their correct primary vertex. An ultimate concept is to develop 4D real-time tracking system for a fast trigger decision [106] and to exploit 5D imaging reconstruction approach, if space-point, picosecond-time and energy information are available at each point along the track.

Several types of technologies are considered for "picosecond-timing frontier": ionization detectors (gas-based devices or silicon detectors), and light-based devices (scintillating crystals coupled to SiPMs, Cherenkov absorbers coupled to photodetectors with amplification, or vacuum devices). The state-of-the-art review of timing studies and summary of the most recent measurements, many from laboratory tests with single pixels, can found in Figure 10 [107,108]. MCP principal use as a timing detector is achieved when coupled to a quartz radiator for Cherenkov light detection. There is a significant progress in MCP-PMT aging properties and rate capability, however, the biggest drawback remains its cost. Typically, the older MCP-PMTs show stable operation up to ~200-300 kHz/cm$^2$ flux of single photons at a gain of $10^6$, while some of the recent ones can push rates up to ~10 MHz/cm$^2$. Recent tests demonstrate remarkable progress of Hamamatsu atomic layer deposition (ALD)-coated MCP-PMT which can be stable up to ~20 C/cm$^2$. An ultimate timing resolution of MCP-PMT has been pushed to ~3.8 ps in beam tests using single-pixel devices. It has been demonstrated in Babar that a suitably polished quartz bar can be used as both a radiator and a minimally-distorting light guide allowing the placement of the light detection system outside of the active volume. Evolution of this technique has been developed for the "Time Of Propagation" (TOP) counter in Belle II, in which both the position and arrival time of Cherenkov light is measured with high accuracy (80 ps) [109], and for the "Time Of internally Reflected CHerenkov light detector" (TORCH) time-of-flight system for LHCb upgrade, with a proposed resolution close to 70 ps per photon resulting in an effective time resolution of 15 ps per track. Panda Endcap DIRC has many similar featured to the LHCb TORCH. Some representative examples of detector technologies used (or proposed) for a precision picosecond-timing frontier detectors are shown in Figure 11.

| Detector | Experiment or beam test | Maximum rate | Maximum anode charge dose | Timing resolution | Ref. |
|---|---|---|---|---|---|
| MRPC presently | ALICE | ~500 Hz/cm$^2$ *** (tracks) | - | ~60 ps/track (present)*** | [4] |
| MRPC after upgrade | ALICE | Plan: ~50 kHz/cm$^2$ ** (tracks) | - | Plan: ~20 ps/track | [4] |
| MCP-PMT | Beam test | - | - | < 10 ps/track * | [7,8,9] |
| MCP-PMT | Laser test | - | - | ~27 ps/photon * | [14] |
| MCP-PMT | PANDA Barrel test | 10 MHz/cm$^2$ * (laser) | ~20 C/cm$^2$ * | - | [11] |
| MCP-PMT | Panda Endcap | ~1 MHz/cm$^2$ ** (photons) | - | - | [28] |
| MCP-PMT | TORCH test | - | 3-4 C/cm$^2$ * | ~90 ps/photon * | [27] |
| MCP-PMT | TORCH | 10-40 MHz/cm$^2$ ** (photons) | 5 C/cm$^2$ ** | ~70 ps/photon ** | [24-27] |
| MCP-PMT | Belle-II | < 4MHz/MCP *** (photons) | - | 80-120 ps/photon*** | [23] |
| Low gain AD | ATLAS test | ~40 MHz/cm$^2$ ** (tracks) | - | ~ 34 ps/track/single sensor * | [34,35] |
| Medium gain AD | Beam test | - | - | < 18 ps/track * | [39] |
| Si PIN diode (no gain) | Beam test (electrons) | - | - | ~23 ps/32 GeV e$^-$ | [8] |
| SiPMT (high gain) | Beam test – quartz rad. | - | < 10$^{10}$ neutrons/cm$^2$ | ~ 13 ps/track * | [8] |
| SiPMT (high gain) | Beam test - scint. tiles | - | < 10$^{10}$ neutrons/cm$^2$ | < 75 ps/track * | [41] |
| Diamond (no gain) | TOTEM | ~3 MHz/cm$^2$ * (tracks) | - | ~ 90 ps/track/single sensor * | [36] |
| Micromegas | Beam test | ~100 Hz/cm$^2$ * (tracks) | - | ~24 ps/track * | [31,32,40] |
| Micromegas | Laser test | ~50 kHz/cm$^2$ * (laser test) | - | ~76 ps/photon * | [31,32,40] |

\* Measured in a test
\*\* Expect in the final experiment
\*\*\* Status of the present experiment

**Figure 10.** Summary of different detector concepts, exploiting picosecond-timing resolution, along with their maximum rate capability and anode charge doses, obtained in beam or laser tests (by J. Va'vra [107]).

Various R&D efforts are ongoing to enlarge the area of the MCP system. A particularly promising one is the LAPPD (Large Area Picosecond Photodector) project [110-112] intended as the direct replacement for the PMT. It targets the development of large-area systems to measure the time-of-arrival of relativistic particles with (ultimately) 1 ps resolution. The rationale for the LAPPD project was to move away from lead-glass MCP's in which the very soft glass was impregnated with hydrogen during (the 80-step!) processing, to a hard glass with the secondary-emitting layer being a hard pure substance, in this case MgO. In addition, the ALD-process was used to simplify the manufacturing of MCPs, to allow tuning of its amplification properties independent of the substrate, and to eliminate the ion feedback destroying the photocathode and limiting the lifetime. In the LAPPD detector, the photoelectrons are accelerated across a proximity gap and then amplified in the MCPs; the resulting electron cloud is detected on anode strips or pads to determine the hit coordinates and the arrival time of the event. Earlier tests demonstrated that the system is radiation tolerant up to at least 7 C cm/$^2$, i.e. more than the conventional MCP, reaches gains of $2 \times 10^7$, has a time resolution for single-photoelectrons below 60 ps and a differential time resolution of below 5 ps for large signals, with an extrapolated number below 2 ps as N/S approaches zero [113], and a spatial resolution of ~ 300 μm using charge sharing [114]. Looking towards the next generation of LAPPD, nanocomposite materials can be tuned to optimize the electrical properties of the MCP. Incom Inc. company has started to produce LAPPDs with RF strip-line anode readout at a rate of a few per month achieving both 10$^7$ gain and a photocathode efficiency of 24%. The US provisional Patent was submitted for a batch production using « air-transfer » technique with a potential to achieve production capacity of 100's of modules/week [115]. If equipped with dedicated custom-made ASICs, LAPPD will have several very interesting potential applications: precision time-of-flight information for a hadron collider, event time measurements in LAr TPC-based neutrino detectors, large water Cherenkov detectors and, beyond HEP, time-of-flight positron emission tomography [116].

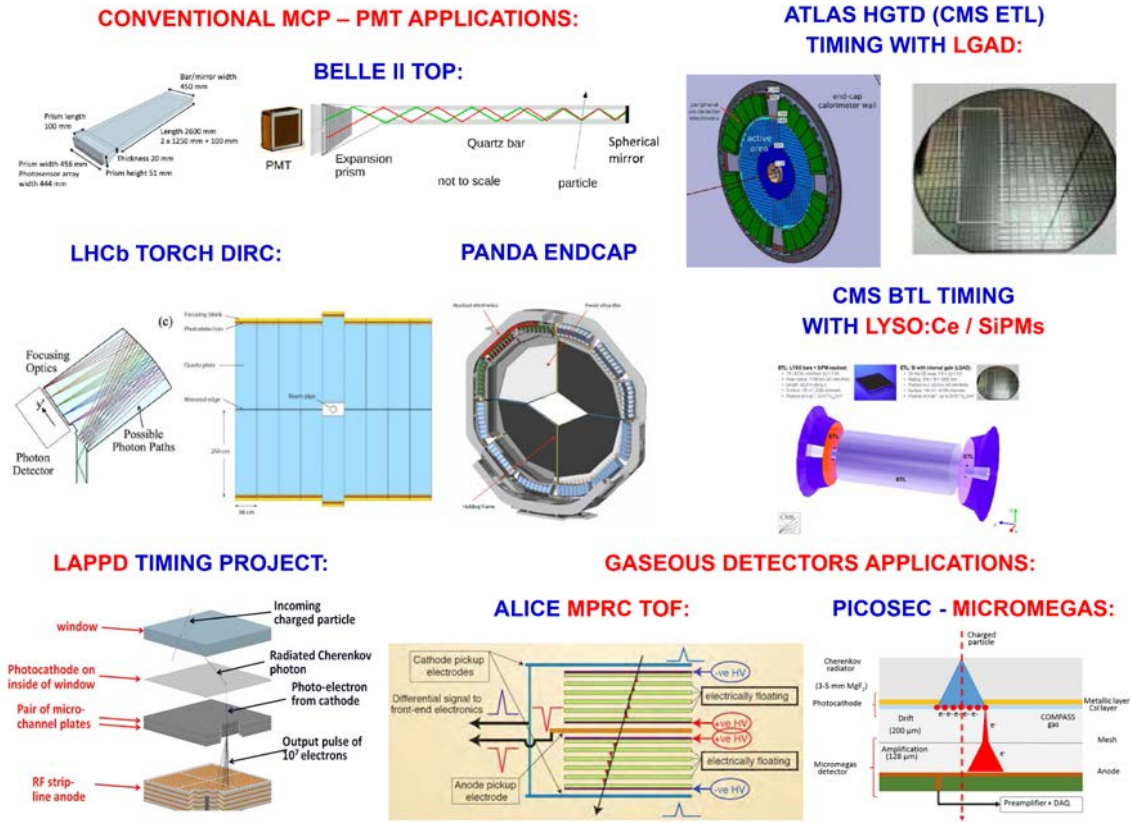

**Figure 11.** Examples of detector technologies used (or proposed) for a precision picosecond-timing concept and some of their applications in the (present/future) large-scale systems.

For the HL-LHC upgrade, both ATLAS and CMS are planning to install new generation of timing detectors between the tracker and the calorimeter, aiming for a time resolution of 30 ps for MIPs. The ATLAS High-Granularity Timing Detector (HGTD) [117] and the CMS Endcap Timing Layer (ETL) [118] are Ultra-Fast Silicon Detectors (UFSD), based on Low-Gain Avalanche Detectors (LGAD) sensors, which will be exposed to fluences of several $10^{15}$ $n_{eq}/cm^2$. Developed jointly with the CERN-RD50 collaboration, LGAD are n-on-p sensors with moderate internal charge multiplication (5-70) due to the presence of a high-field region below the junction, formed using a deep "reach through" implant. Aiming for an excellent position and timing resolution (~ 10 ps and ~ 10 μm) with GHz counting capabilities to perform 4D tracking, LGADs still suffer from low fill factor and have a moderate radiation hardness. Recently, 35 ps have been obtained in 50 μm devices at a gain of 20, while studies to improve radiation hardness focus on optimization of the implant that defines the gain layer. LGADs are currently being produced by 3 foundries (CNM, FBK, and HPK) and their exact implementation still requires some R&D. Obviously, sensors characterized by such a time resolution represent a very attractive option for PID and TOF applications. As a viable alternative to UFSD, LHCb is developing a time-tracking O(100 ps) device, based on 3D trench Si-sensors with a more uniform field/charge collection, and a goal to withstand higher fluences of $10^{16}$-$10^{17}$ $n_{eq}/cm^2$ [119]. Good timing resolution (~100 ps) can be also achieved with ultra-pure mono-crystaline sCVD diamond sensors. This concept can be used as a preshower for fast-timing calorimeters and future LHC pp-diffraction scattering experiments.

While the radiation levels in the endcap regions of ATLAS and CMS make silicon detectors such as LGAD sensors the only viable solution, fast scintillating crystals (LYSO:Ce) with SiPM readout, cheaper than all-silicon solution, are possible for the CMS barrel system [118]. In principle, the idea of using scintillating crystals coupled with photodetectors with high gain for a picosecond-timing detectors comes from the TOF-PET concept. LYSO is well suited due to its high light yield and fast signal rise time; test beam measurements have demonstrated that sub-30 ps resolution can be achieved with single crystals and SiPM readout. The CMS barrel timing layer (BTL) will cover an area of approximately 36 $m^2$ (~250,000 LYSO tiles with a size of ~ 12 x 12 $mm^2$ readout by ~ 332,000 SiPMs with an active area of ~ $4 \times 4$ $mm^2$), aiming for a MIP timing resolution of ~ 30 (60) ps at the beginning (end) of the lifetime. The detector has to fit within a 25 mm thick envelope between the CMS tracker and the electromagnetic calorimeter. The high radiation levels (~$10^{14}$ $n_{eq}/cm^2$) requires operation at -30°C using $CO_2$ cooling; same as for the CMS Endcap High Granularity Calorimeter, discussed in Section 6.

Gaseous detectors may also have a chance of achieving picosecond timing precision. Existing ALICE TOF detector, based on multi-gap resistive plate chambers (MRPC) and covering an area of ~160 $m^2$, has achieved a time resolution of ~ 60 ps. New studies on MPRCs with 20 gas gaps using a low-resistivity 400 μm-thick glass aims to achieve ~ 20 ps resolution per MIP for an upgraded detector [120]. An interesting development for the future TOF applications is represented by the implementation of neural networks for MRPC time reconstruction, instead of the standard tile-over-threshold (TOT) technique [121]. In the context of the PICOSEC-MM project, a two-stage amplification Micromegas coupled to a Cherenkov radiator coated with a semitransparent CsI photocathode achieved a time resolution of 24 ps per MIP for a 1 cm prototype [122]. Further R&D is ongoing to develop readout electronics, verify the scalability to target larger sizes and photocathode (PC) robustness at large particle rates. New DLC-like PCs are under investigation; so far, their QE is a few times lower than of CsI.

Regardless of the type of solution chosen, optimizing the performance of a large-scale system to achieve a time resolution of better than 50 ps is not a trivial task and requires a careful optimization of all parameters. In addition to the sensor-level uncertainty, which comes from many sources, such as signal jitter, detector inhomogeneity, Landau fluctuations, TDC bin size, a baseline clock distribution and monitoring system across different large-scale subdetectors has to be maintained (jitter smaller than ~ 15 ps rms). Once detector technology choice is made, full production of large-scale detectors requires a careful quality control.

## 5. Advanced Concepts in Particle Identification and Photon Detectors

Particle identification methods over a large part of the phase space and particle types have become an indispensable tool in heavy flavour physics, heavy-ion and nuclear physics experiments [123-126]. The choice of the PID detector is driven primarily by the physics performance and many clever techniques have been developed. State-of-the-art PID system can be illustrated by means of the ALICE experiment, which rely on three legs with different ranges of momentum sensitivity: dE/dx, TOF, Cherenkov and Transition Radiation (TRD) detection. Figure 12 (left) illustrates schematically the momentum ranges in which various PID methods are exploited in ALICE [127]. Hadrons are identified by their mass, which is, in turn, determined by combining the measurements of momentum and velocity, the latter is being derived from Cherenkov angle (and/or number of photons), ionization losses or time-of-flight information. Electron-hadron separation is exploited in a hadron-rich environment such as a heavy-ion colliders RHIC or FAIR, e.g. Cherenkov threshold counter called Hadron Blind Detector (HBD) with GEM readout was used in the PHENIX experiment. It is also worth mentioning that e/π separation has been effectively used in the ATLAS transition radiation tracker. In general, since the PID system is confined between the tracking and the calorimetry system, the specific technology is constrained by the accelerator environment (event rates, triggering) and by the experimental set-up (material budget, size and space requirements, accessibility, compatibility with other detector subsystems, coverage).

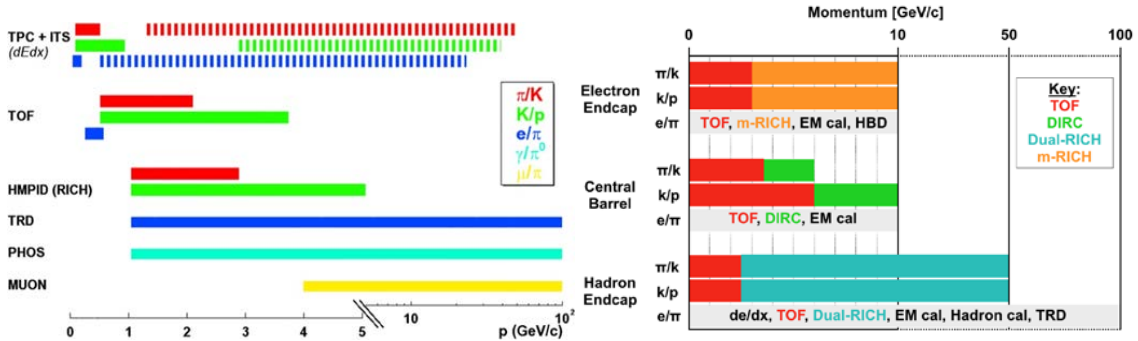

**Figure 12.** (left) Illustration of momentum coverage for various PID methods (dE/dx, TOF, RICH, TRD) in the ALICE experiment; (right) Several detectors are required for a combined PID system at the future Electron-Ion Collider (EIC). The chart represents only a general example and the final system will depend on several factors (by H. Montgomery)

Cherenkov counters became a practical technique for particle detection following the invention and development of PMTs over 70 years ago. They may be classified as either threshold or imaging types. In the former case, photons are counted in order to separate particles below and above threshold. Imaging counters may be used to track particles as well as to identify them, by measuring the ring-correlated angles of emission of the individual photons. In a Ring Imaging Cherenkov Detector (RICH), Cherenkov radiation emitted in a transparent dielectric medium (radiator), whose refractive index is appropriate for the range of particle momentum being studied, is transmitted through an optics, which could be either focusing with a spherical (or parabolic) mirror or not focusing (proximity-focusing), onto a photon detector that converts photons into photoelectrons with a high spatial and time resolution [128]. Good segmentation in space may be needed to obtain adequate angular resolution; however, many costs scale with the number of pixels and must be controlled. Over the years, an admirable workmanship in the design and construction of radiators with many different techniques has been developed: gaseous, solid-state, liquid, aerogel, at low or high temperatures. Optical components in Cherenkov imaging detectors are becoming increasingly important and the requirements now include mirrors, lenses and systems to control and monitor the alignment of huge optical arrangements. With the present LHCb RICH, single-photon Cherenkov resolution of 0.7 mrad has been achieved. In well-focused R&D, a better than 0.2 mrad resolution could be demonstrated, even in presence of high event multiplicities and complex topologies [26]. A proximity focusing aerogel RICH detector, manufactured as a multilayer stack with a non-homogeneous aerogel radiator, has driven the design of the ARICH for Belle-II [129], Forward RICH for PANDA detector [130], and FARICH for the Super Charm-Tau factory project.

An advanced TOF detector, in which particles produce Cherenkov radiation in the entrance window of a PMT-MCP, has been proposed to advance measurements of particle velocities at a future hadron or lepton collider. Such system could potentially achieve timing resolution of the order of O (10 ps) [131]. Using the same idea, to detect Cherenkov photons emitted in the PMT entrance window, a generic R&D study of a detector, combining proximity focused RICH with aerogel, has been carried out recently. Such a TOF device with a ~ 40 ps resolution could allow to extend PID capabilities below the threshold in aerogel [126].

A very peculiar way to reconstruct Cherenkov images was pioneered in the BABAR experiment; the novel device, called DIRC (Detection of Internally Reflected Cherenkov light) exploits the photons totally reflected in a long quartz bar rather than those refracted out from the radiating medium as in a classical RICH detector [132]. The DIRC-derived detectors are largely based on the exceptional time resolution of a few tens of picoseconds that can be achieved with MCP-PMTs. A wide range of modern developments are now underway using the DIRC technique, including the Time-of-Propagation (TOP) counter at Belle II [109], a number of FDIRC designs for GLUEX [133], PANDA barrel and endcap DIRCs [134]; and the TOF counter proposed for the LHCb, called TORCH. A sophisticated system of Cherenkov detectors, consisting of a compact focusing DIRC in the Central Barrel, aerogel modular RICH in the Electron Endcap and Dual-RICH in the Hadron Endcap is been designed for a large range of solid angles and particle momenta for the future EIC collider [135]. The final configuration will depend on several factors; a chart representing an example solution is given in Figure 12 (right).

Fast single photon detectors are essential components in many particle identification (and calorimetry) systems. There are several distinct types: vacuum, solid-state, gaseous and superconducting photon detectors, and a continuous push for higher efficiency, dynamic range, radiation hardness, lower noise and improved time resolution.

Many different vacuum photon detectors have been used for RICH, such as dynode-based PMTs, micro-channel plate PMTs (MCP-PMTs), hybrid photodetectors (HPD), etc … The quality of the photon detector is expressed in terms of "photon detection efficiency" (PDE), which is defined as the product of the quantum efficiency (QE) of the active area, the fill factor (ratio of sensitive to total areas) and the probability that an incoming photo-electron undergoes secondary multiplication and is eventually detected. A wide variety of photocathodes, with different techniques for obtaining gain, are available that are sensitive to wavelengths from the UV cutoff of the window material (LiF cuts off around 100 nm) to the near IR. Both "transmission" and "reflection"-type PMTs has been the workhorse of photo-detection for over fifty years, providing robust, low noise, detection of single photons with nanosecond timing. Solutions have been developed for operation at cryogenic temperatures and in high magnetic fields; however these extreme environments remain challenging. The multi-anode PMT (MaPMT) consists of a square array of metal dynode chains incorporated into a single vacuum tube with a very little cross-talk between individual channels. They will replace HPDs, as a part of the major upgrade for both LHCb RICH counters, to prepare for a higher event occupancy during the LHC Run III [136]. MCP-PMTs, a photomultiplier tube with two microchannel plates instead of the dynode structure, is considerably faster (~20 ps) than a standard PMT and can tolerate random magnetic fields up to 0.1 T and axial fields exceeding 1 T. They were used for the first time on a large scale in Belle II TOP. A major breakthrough has been achieved in mitigation of the PC aging, due to ions liberated by secondary electrons from the surface of the MCP pores, by employing a novel production technique, where the secondary emitter is deposited by means of atomic layer deposition (ALD). Such a new ALD-PMTs will replace existing devices in Belle II TOP in 2021 [109]. HPDs combine the sensitivity of a vacuum PMT with the excellent spatial and energy resolutions of a Si-sensor. They achieve ~10 ps resolution, but function in an axial magnetic field only if they are proximity-focusing type. In addition to LHCb RICH, HPDs have been also used on a large scale in the original hadronic barrel and endcap CMS calorimeters, but are in the process now of being replaced by the SiPMs. The motivation comes from two key advantages of SiPMs over the HPDs, namely, an approximately three times higher PDE, which directly reduces the impact of the light yield degradation of the scintillator system, and a much smaller physical size, which allows to install a larger number of photon sensors, providing a finer longitudinal segmentation of the calorimeter.

Gaseous detectors are not commercially available but represent the most cost-effective solution to cover very large areas with photosensitive elements; they allow minimal material budget and their operation is compatible with the presence of a magnetic field [137]. Gaseous (TMAE, TEA) and solid (CsI) photocathodes have been usually employed. Today, the largest CsI-MWPC based RICH counter is the ALICE High Momentum Particle Identification Detector (HMPID), covering an area of 10 $m^2$. Conceived with the aim to overcome the limitations of MWPCs, MPGD-based gaseous photomultipliers with semi-transparent or reflective CsI-PC allow to minimize the ion and photon feedback to the PC, to preserve it from aging, and to avoid secondary effects causing electrical instability. To achieve a further IBF suppression an alternative hybrid architecture, combining MM and THGEM, has been developed and installed for the COMPASS RICH-1 upgrade [138]. As main drawbacks, all MPGD-based photon detectors are too slow for focusing DIRC applications and their response is limited to the UV region. Therefore, they suffer from larger chromatic dispersion compared to vacuum photomultipliers equipped with bialkali-PC and their performance at high luminosities might be severely limited by the photocathode aging. Hydrogenated nano-diamond crystals are being studied these days for a compact RICH at the EIC.

In the solid-state photon detectors realm, these include visible-light photon counters (VLPC), photodiodes with no internal amplification, avalanche photodiodes (APD or HAPD) and silicon-photomultipliers (SiPMs, MPPCs). The latter are formed by a large array of very small avalanche photodiodes operated in Geiger mode and reaching a gain of $10^6$. They have several advantages: insensitivity to the high magnetic fields, lower operation voltage, and less material in comparison with conventional PMTs. Due to their small dimensions, they allow compact, light and robust mechanical designs. Significant progress in understanding of SiPM physics was achieved during the last five years. Their cross-talk and after-pulse effects are now reduced to a few percent level and the PDE is about ~50-65% for blue/green light, which is significantly higher than that of commonly used PMTs. However, the major drawbacks, such as very high dark count rate and moderate radiation hardness, have so far prevented their widespread use in the RICH detector, where single photon detection is required at low noise. Because the Cherenkov light is prompt, the problem of high dark count rate can in principle be reduced by using a narrow time window (~3 ns) for signal collection. In addition, it is possible to further increase the signal-to-noise ratio by using light collection systems. Several tests have shown that SiPMs can indeed be used as a promising candidate for the detector of Cherenkov photons [139].

In general, many advances are now being made with devices based on superconductivity, especially those with applications for cosmology (not discussed in this paper). Examples are transition edge sensor, kinetic inductance detector, CMOS-based single photon quantum-dot device (with a direct coupling of quantum dots to amplification circuit), single photon detector based on carbon nanotubes, etc. The ultimate challenge for the modern developments is a detection of optical signals at a quantum level – resolving arrival time and spatial location of individual photons. Development of the 3D SiPMs, with low power consumption and ~ 10 ps timing resolution, might be a breakthrough in photon detection comparable with the development of the Quanta Image Sensors (QIS) [140].

## 6. Advanced Concepts in Calorimetry

The story of modern calorimetry is a textbook example of physics research driving the development of an experimental method [141,142]. Present and future challenges in calorimeters are closely linked to all aspects of ultimate exploitation of the particle flow (PFlow) technique [143] or the dual-readout calorimetry approach [144]. They can be further classified according to their construction technique into sampling and homogeneous calorimeters. Homogeneous electromagnetic calorimeters (ECAL), where is entire volume contributes to the signal, may be built exploiting three primary signal collection mechanisms: scintillation light from inorganic heavy (high-Z) crystals (NaI, CsI, LSO, BGO, BSO, $BaF_2$, $PbWO_4$, LYSO, LuAG, YAG, GAGG, etc …), non-scintillating Cherenkov light (lead glass and $PbF_2$ radiators) and ionization noble liquids (Ar, Kr and Xe). A sampling calorimeter, which is a practical solution for higher energy ranges and for hadronic calorimetry (HCAL), consists of absorber of high density (Pb, W, Fe, Cu or depleted U) sandwiched or (threaded) with an active material that generates signal (scintillator, ionizing noble liquid, gaseous or semiconductor detector). Research and development in the field has resulted in the use of many different technologies, depending on the application: liquid noble gas, crystals, scintillator-based sampling, and silicon-based sampling calorimeters.

Noble liquid calorimeters have long been used in sampling electromagnetic calorimetry (ECAL). Nowadays, LAr application in ATLAS [145] derives from its radiation hardness, good energy resolution, high rate capability and pile-up suppression. The granularity of noble liquid calorimeters can easily be adjusted to the needs by finely segmented read-out electrodes (multi-layer PCBs). New studies of LAr properties, under high ionization rates, cover space charge build-up, initial and bulk recombination, surface charge accumulation and the role of impurities, are essential. High granularity LAr calorimeters will remain a major technique at the Energy Frontier, mainly due to the intrinsic radiation hardness, 3D imaging capability and good timing resolution, which is based on the fast signal rise-time and homogeneous active material. Hence, LAr is part of the reference design of FCC-hh and was also recently considered for FCC-ee detector.

Homogeneous crystals offer the best possible energy resolution for electromagnetic showers. Light yield, radiation length, decay time, and radiation hardness vary enormously among the different materials and are extremely sensitive not only to voluntarily introduced dopants, such as Tallium in CsI, but also to any undesired impurity during the high temperature crystal growth process. With high density (7.1 $g/cm^3$), bright light yield (200 times of $PbWO_4$) and fast decay time (30 to 40 ns), LYSO crystal is a natural choice, but it remains prohibitively expensive for most large scale applications. Therefore, cost-effective crystals, such as CsI, $BaF_2$ and $PbWO_4$, are often used. CsI(Tl) is well matched (light yield and resolution) to the requirements of low-rate, low-radiation e+e− flavour factories, e.g. Belle II ECAL [146], and an undoped CsI calorimeter with SiPMs readout will be installed in the Mu2e experimental hall in 2021 [147]. Among all existing crystal calorimeters, the CMS lead tungstate, consisting of 75,848 crystals of 11 $m^3$, is the largest one, but it faces a challenge of severe radiation environment. Significant losses of light output (> 50% after LHC Run II) have been observed in crystals at large rapidity caused by both ionization dose and hadrons. Thanks to the CMS ECAL laser monitoring system, residual energy scale corrections are kept at ~ few % level without dependence on instantaneous luminosity [148]. Examples of crystal calorimeters currently under construction are: a PbWO$_4$ calorimeter for PANDA at FAIR [7], a LYSO calorimeter for COMET at JPARC, and a PbF$_2$ calorimeter for the g-2 experiment at Fermilab. A new concept called Dual-Gate Calorimetry has emerged, where the independent and simultaneous detection of Cherenkov and scintillation light enables the measurement of the electromagnetic fraction of hadron shower event-by-event. This principle has been successfully tested on copper fiber calorimeters, but it can be in principle also applied to homogeneous crystals, e.g. BGO or BSO. Development of the cost-effective UV-transparent and inorganic scintillators, including crystals and glasses, may achieve unprecedented jet-mass resolution with dual-gate readout for a homogeneous hadron calorimeter concept at a future circular lepton collider. In particular, the FCC-ee experiment is investigating an option of the dual readout fibre calorimeter.

Scintillator-based calorimeters are of interest for many applications at future Lepton Colliders, due to its cost-effectiveness and moderate radiation hardness. Homogeneous crystal, shashlik-type and spaghetti-type calorimeters are considered for the LHCb ECAL Upgrade in a more challenging radiation environments (neutron flux is up to $3\times10^{15}$ neutrons/cm$^2$ and a Total Ionizing Doses (TID) of 3 MGy [149]). All such calorimeters will profit from the excellent timing resolution to suppress pile-up; the requirement is ~ 20-50 ps. An advanced machine learning technique (XGBoost) was developed to improve spatial reconstruction and time of arrival evaluation for the LHCb ECAL Upgrade Phase II [150]. Due to the relatively modest radiation environment, technology similar to the ATLAS tile calorimeter (iron plates as an absorber with plastic scintillating tiles and wavelength shifting fibres) [151,152] is under investigation for the FCC-hh hadronic barrel calorimeter.

All known inorganic scintillators suffer from damage induced by ionization dose as well as charged and neutral hadrons. Therefore, R&D on multi-purpose inorganic scintillation materials is in full swing [153,154]; many aspects have also been inherited from strong efforts of the Crystal Clear Collaboration (CERN-RD18: study of new fast and radiation hard scintillators for calorimetry at LHC). There are three possible radiation damage effects in crystal scintillators: scintillation mechanism damage, radiation induced phosphorescence (afterglow) and radiation induced absorption (color centers). LYSO, BaF$_2$ crystals and LuAG ceramics show excellent radiation hardness up to 340 Mrad, $1\times10^{15}$ protons/cm$^2$ and $3.6 \times10^{15}$ neutrons/cm$^2$, which is adequate for HL-LHC. Following these studies, LYSO crystal was proposed for the CMS BTL detector at HL-LHC. An important question concerns the radiation hardness of potentially interesting radiation-hard crystals like YAG and GAGG, at low temperatures (-30°C to -40°C), as required by the operation of irradiated SiPM photodetectors. GAGG crystals are also considered as the best option for a novel radiation-hard ECAL planned for the LHCb Upgrade II.

Today, calorimetry is the dominant area of the SiPM use in particle physics [101]. In addition to the common application of reading out plastic scintillators, SiPMs are well-suited for crystals in electromagnetic calorimeters. SiPMs could also enable substantial improvement in dual-readout calorimetry, based on scintillating and clear plastic fibers embedded in absorber structures. One of the challenges is the production of devices that are themselves sensitive in the VUV, limiting the need for secondary wavelength shifters. The design choices for the systems are driven by specific properties, limitations and requirements of SiPMs. Examples are the cold operation at -30C or below in the LHC experiments to keep the radiation damage-induced dark rate at tolerable levels, precise coupling units to match fibers guiding the light to the SiPM to the sensor active area. SiPMs find natural application in detectors that already require cooling like the noble liquid TPCs (see Section 9). An optimization of the SiPM design for operation in high radiation environment is ongoing [155]; recent studies revealed that the HPK and FBK SiPMs are still operational up to fluences of $\simeq 2 \times 10^{14}$ $n_{eq}/cm^2$ at -37C. The main limitation comes from the high power dissipation in irradiated sensors, caused by significant dark current increase [156]. Figure 13 shown example of the SiPM (FBK, W9C) performance before and after irradiation.

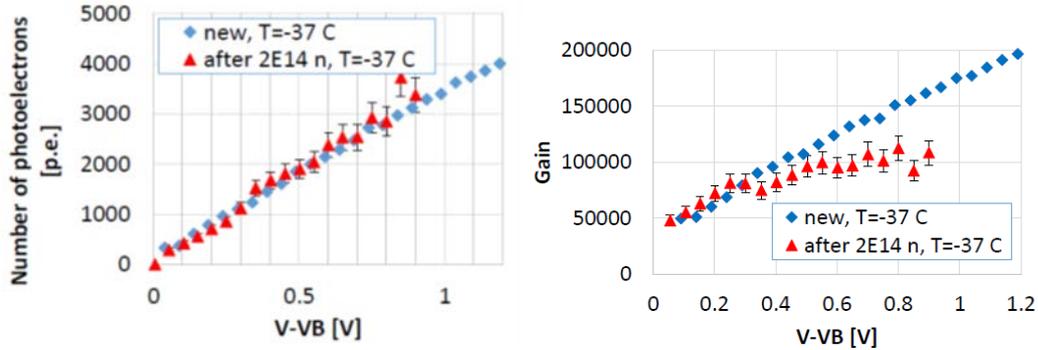

**Figure 13.** Example of the FBK W9C SiPM performance, which is still operational after irradiation dose of $2 \times 10^{14}$ $n_{eq}/cm^2$ at -37C. The main limitation in improving S/N ratio is high power dissipation in SiPM that increases p-n junction temperature and causes $V_B$ shift and additional dark current increase [156].

R&D in particle flow calorimetry is a major "paradigm shift" for high-resolution imaging calorimeters. It has been primarily an electron-positron Linear Collider driven effort, but nowadays is highly relevant for HL-LHC upgrades and future Energy Frontier colliders. The idea is to measure the energy of all the particles in a jet, using the tracker for the charged particles, ECAL for prompt photons, and HCAL to capture the neutral hadrons. Due to this, PFlow has led to calorimeter designs, as part of a complex system of inter-connected detectors rather than as a stand-alone device. The PandoraPFA algorithm was developed to study shower substructure and optimize PFlow performance for the ILC and, later, was successfully extended to CLIC energies, where the particle flow becomes more of an energy flow [157]. The concept of digital ECAL brings potential to measure the number of secondary interactions in a particle shower, which follows a Gaussian distribution, and removes the Landau fluctuations associated with the energy deposited in thin layers of material. Figure 14 summarizes different types of imaging calorimeters developed for Linear Colliders; most projects were advanced within the CALICE collaboration [158]. The different calorimeter R&D groups followed more or less the same (impressive) path that started with sensor and readout development, followed by building a small-scale prototype and performing beam tests for detector characterization, then followed by the construction and extensive beam tests of a large "physics prototype", that serves as a proof-of-principle for a given calorimeter concept, and finally proceed to the construction of "full technological prototype" that addresses all aspects (power pulsing, compact mechanical design, embedded electronics, scalability and assembly, calibration approaches and tools, etc) in a realistic detector. A full hadronic technological prototype (AHCAL) with a volume of approximately 1 m$^3$, based on 3×3 cm$^2$ scintillator tiles and ~ 7500 SiPMs integrated with the embedded read-out electronics, was constructed making use of automatized testing and assembly procedures. The AHCAL prototype in 2006 was the first large-scale application of SiPMs in high energy physics [159]. The time structure of hadronic showers was studied at the CERN test-beam with two technological prototypes: AHCAL and semi-digital hadronic calorimeter with RPC readout (SDHCAL) [160]. The comparison of detailed detector simulations with the data revealed that the time structure is generally quite well modelled in the steel absorber, while, the tungsten data was only reproduced by models with a dedicated treatment of low-energy neutrons. A new particle identification method based on the BDT MVA technique, using a topological shape of events, has been developed to distinguish between the hadronic and the electromagnetic showers [161].

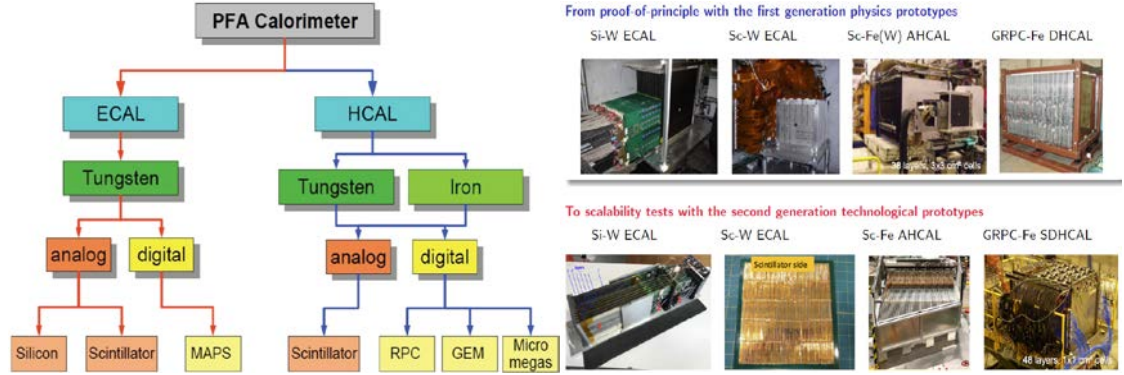

**Figure 14.** (left) Overview of a rich program exploring the full spectrum of imaging calorimeter technologies, driven by the CALICE Collaboration, which are being considered for a future Linear Collider; (right) CALICE collaboration pioneered developments of highly granular calorimetry concept from generic R&D (since 2005) to large-scale "technological prototypes" [161].

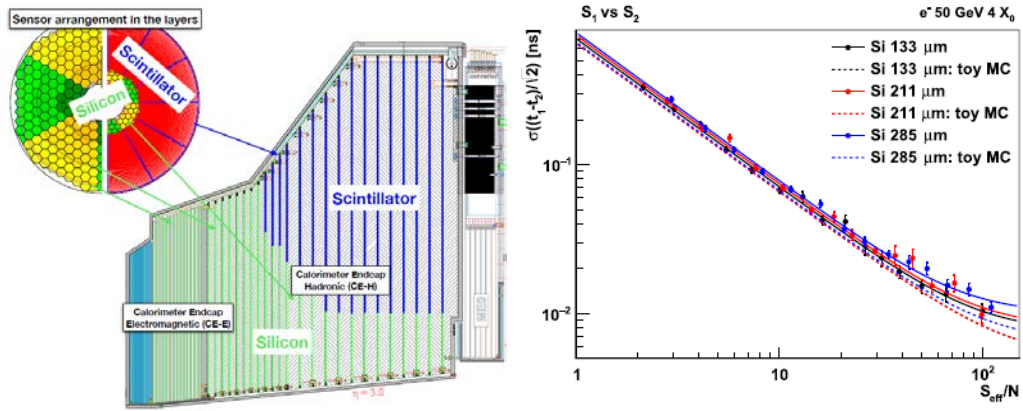

**Figure 15.** (left) Longitudinal cross section of the CMS High Granularity Endcap Calorimeter (HGCAL). The green region to the lower left is instrumented with silicon detectors and the blue region to the upper right with scintillator tiles; (right) Timing resolution based on silicon-sensors as the function of signal-to-noise ratio; 30 ps could be achieved for S/N > 20 [163].

In the very forward region of the ILC, two ECAL calorimeters, Lumical and Beamcal, are foreseen to provide instantaneous and integrated luminosity measurement and also used for the beam tuning/diagnostics [162]. Finely segmented radiation-hard Si (LumiCal) or GaAs(BeamCal) sensor planes with a dedicated fast readout are being developed by the FCAL collaboration.

A long-standing concept for the ILC, a highly granular endcap imaging calorimeter (HGCAL) for the CMS Phase-II upgrade will bring another change in paradigm. It effectively 'tracks' particles inside the calorimeter enhancing the PFlow concept. The longitudinal cross section of the HGCAL is displayed in Figure 15 (left). It will consist of silicon pad sensors (≈0.5–1.2 cm$^2$) of different thicknesses (120 to 300 μm, thinner at smaller radius where radiation levels are higher) for the ECAL (28 Si-layers) and front section of HCAL (24 Si-layers) and highly-segmented plastic scintillator tiles (≈4–30 cm$^2$) with SiPM readout for remainder of the HCAL; the latter is to reduce the overall cost of the system [163]. In total, 600 m$^2$ of silicon sensors and 500 m$^2$ of scintillator elements with a total of 400,000 SiPMs will be used. The full calorimeter will be situated in the same cold volume at -30∘C. The Si sensors of the HGCAL will be exposed to fluence ranging from about $2\times10^{14}$ up to about $10^{16}$ n$_{eq}$/cm$^2$; the transition between the silicon and scintillator parts is defined by the radiation damage and the corresponding loss of the scintillator light yield, and the high dark count rates of the photon sensors through the full HL-LHC operational lifetime of 3000 fb$^{-1}$. The main differences between CMS HGCAL and ILC/CLIC reside in the readout timing and the power pulsing, which enables a larger effective density of the calorimeter and more compact particle showers. A silicon-tungsten calorimeter is also proposed for FCC-ee and a promising option for a FCC-hh calorimeter, in areas where the radiation level compatible with the technology.

There has been another major "paradigm shift", which is to perform picosecond-timing measurements in large-scale calorimeters, as the technology matures. Work is ongoing to demonstrate that ultra-fast timing can be used to resolve the development of hadron showers and therefore improve the resolution of hadron calorimeters. Results from beam tests have shown that the timing resolution obtained with Si-sensors does not vary significantly with a thickness as a function of S/N [164]. Figure 15 (right) shows that 30 ps time resolution for S/N > 20 can be achieved for charged particle detection. The principle sources of uncertainties are Landau fluctuations in the signal, slewing in the preamplifier, and inter-channel variations in the clock distribution system. The time of arrival of electromagnetic showers in calorimeters could be measured with significantly better precision using the large number of synchronous particles in a shower, and showers spreading over a large number of independent sensors. Multiple samples of the shower with timing measurements in each layer can be combined to improve the time-of-arrival performance. Bright inorganic scintillators and silicon detectors are very promising for the development of high precision timing measurements in calorimeters. Design of low-cost and radiation-tolerant electro-optical transceivers and front-end electronics is therefore critical to obtain a high-precision at system level [31]. Precision timing measurement of photons in ECAL will allow to triangulate and select a set of possible vertices in the Higgs boson two-photon decay, mitigating the pile-up effect at HL-LHC.

## 7. Advanced Concepts in Electronics, Trigger and DAQ

While research has always required state-of-the-art instrumentation in trigger and data acquisition (TDAQ) systems, the demands for the next generation of hadron colliders are the increasingly large local intelligence, integration of advanced electronics and data transmission functionalities (e.g. using Field Programmable Gate Arrays (FPGA) [165]). Rapid development in FPGA processor power and I/O bandwidths are also enabling complex online feature extraction and increased readout flexibility. One of the most important trends that emerged from the LHC designs is the widespread adoption of low-latency, high bandwidth (rad-hard optical link ~ 5 Gb/s today) cluster interconnect applications. Optical data transmission technology offers a very elegant way for readout of modern HEP detectors. Optical links in state-of-the-art detector systems are implemented based on vertical cavity surface emitting lasers (VCSEL), where the information is encoded onto the optical carrier by varying the laser current. Another high-performance optical link, based on silicon photonics and wavelength-division multiplexing (WDM) technology, to cope with ultra-broad bandwidth requirements has been proposed in [166]. An important challenge is to extract the data at high bandwidth from the tracking detectors without adding a prohibitive material budget due to the large number of fiber drivers and the power and cooling they would require. Meanwhile, other readout concepts are being investigated. Free-space optical links, offer some advantages for interlayer communication by removing the problem of radiation damage of the optical fibers, together with opto-electromechanical devices, for an elegant implementation of intelligent front-end interconnects [167]. A feasibility study of the wireless technology integration at 60 GHz into silicon tracker detector has been reported in [168]; it represents a promising upcoming alternative to conventional data transmission.

Another important trend is the progressive replacement of the complex multi-stage trigger systems by a new architecture with a single-level hardware trigger and a large farm of Linux computers to make the final online selection and to reduce Level-1 trigger rate to O(kHz) for permanent storage. An example is an upgraded ATLAS TDAQ baseline architecture with a single-level hardware trigger that features a maximum rate of 1 MHz and 10 µs latency; several reports describing new ATLAS trigger and electronics developments were presented in [169-172]. The ALICE and LHCb experiment will be triggerless (i.e., no hardware trigger level) after the LHC Phase I upgrade. Whether triggerless or not, Higher Level Triggers would have to process at least ten times as much data as the present LHC detectors. This clearly illustrates the trend towards moving more complex algorithmic processing into the online systems. A possible evolution of TDAQ architectures could lead to detectors with extremely deep asynchronous or even virtual pipelines, where data streams from the various detector channels are analysed and indexed in situ quasi-real-time, using intelligent, pattern-driven data organization, and the final selection is operated as a distributed "search for interesting event parts" [173].

Modern technologies allow the integration of significant intelligence at the sensor level and many different R&D lines are being explored, like local hit clustering for strip and pixel detectors, local energy summing for calorimeters, local track-segment finders. As an example, CMS will introduce FPGA-based "track-trigger" concept for the outer tracker at HL-LHC, which will be the first detector providing momentum discrimination at the L1 trigger level [174]. The "stub" concept as well as scheme of $p_T$-modules is schematically shown in Figure 16 (left). The recognition of particles with high transverse momentum is based on modules, which are made of two closely spaced adjacent detectors. The FE-chips at the sensor ends read channels from both sensors and correlate the signals. The readout electronics measure the bending between the crossing points of a particle in these layers. The availability of fully reconstructed high $p_T$-tracks ("direction vectors") before the HLT allows a significant reduction of fake tracks. A concept for an FPGA-based track finder using a fully time-multiplexed architecture was developed, where track candidates are reconstructed using a projective binning algorithm based on the Hough Transform, followed by a combinatorial Kalman Filter. The choice of a 65 nm CMOS technology for readout ASIC made it possible to satisfy power requirement despite the fairly large amount of necessary logic to perform the momentum discrimination and the continuous operation at 40 MHz [175].

The HL-LHC upgrade is by far the single largest ASIC demand for high energy physics this decade. Future ASICs R&D challenges include smaller pixels to resolve tracks in boosted jets, very high hit rates (1-2 GHz/cm$^2$), unprecedented radiation tolerance (10 MGy), much higher output bandwidth, and large IC format with low power consumption in order to instrument large areas while keeping the material budget low. The state-of-the in the front-end-electronics design is represented by the activity of the CERN-RD53 collaboration [30]. Its scope is a joint ATLAS-CMS effort to develop, in 65 nm CMOS, radiation-hard ASICs with two possible aspect ratios of ~50x50 μm$^2$ or 25x100 μm$^2$ (including readout software and prototypes), low threshold (< 1000e−), high hit and trigger rate (thanks to up to four 1.28 Gb/s output links), radiation resistance and serial powering capabilities. ATLAS and CMS are not the only projects studying the feasibility of 65 nm technology in HEP. Other examples include the Belle II experiment, which uses a gigabit data transmitter; this chip features high-speed links running at 1.6 Gbit/s, with a total bandwidth of up to 6.4 Gbit/s, to read out DEPFET modules. Another project developed at CERN is the Lp-GBT, a low-power redesign of the Gigabit Transceiver (GBT) ASIC tailored to implement multipurpose high-speed bidirectional serial links [176].

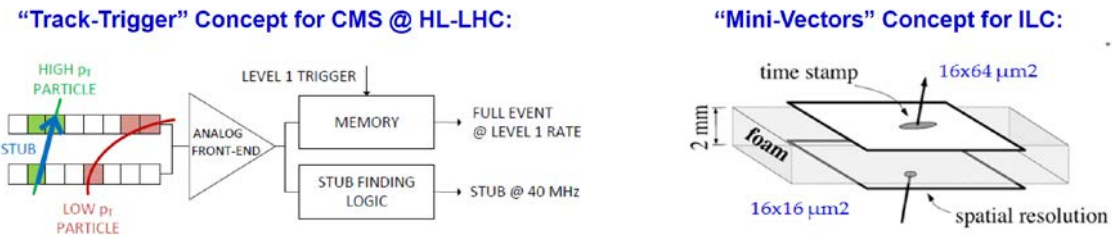

**Figure 16.** (left) A simplified view of the CMS L1 "track-trigger" concept. Coincidence of signals in closely-spaced layers allows to reject low-$p_T$ particles, by using correlation logic implemented in the front-end ASIC; (right) A schematic drawing of the "mini-vector" concept based on double-sided CPS layers for the ILC vertex detector.

Another example of the "intelligent tracker" is a "mini-vector" concept of double-sided CMOS pixel sensor (CPS) layers developed for the ILC vertex detector [56]. The track position and timing may, for instance, be derived from high precision sensor located on one layer side and a faster, less precise, sensor located on the opposite side, thereby assigning to each reconstructed particle trajectory the required spatial and timing resolutions, as shown in Figure 16 (right). The technology of the fast sensors may be CMOS or LGAD. A feasibility study to develop CMOS pixel sensor with on-chip artificial NN, in order to remove low-momenta background particles, was validated by an offline methodology [177]. Another promising R&D direction addresses the realisation of two-tier chips interconnected at the pixel level through industrial micro-bonding techniques. This approach, pioneered with SoI sensors, is getting extended to CPS.

While microelectronics technologies have been key to the success of tracking systems, 3D integrated circuitry (3D-IC) vertical interconnection of multiple layers of CMOS electronics enables viable solutions for the very fine pitch (4 mm), high-density pixelated devices. Wafer thinning allows construction of low mass, high resolution sensors. Vertical vias running through layers of silicon (TSVs) are a key ingredient for 3D-stacking. In addition, TSVs can reach through to the backside of a chip, allowing re-distribution of electrical connections on the chip's backside metal, as it was done for the ATLAS FE-I3 chip wafers. The original motivation of exploring 3D-IC in particle physics was a compact pixel detector for the ILC. An international consortium from 15 institutes was formed that submitted a 3D multi-project run to Global Foundries, through Tezzaron in May 2008, with 3 different ASIC designs, in 130 nm CMOS, to demonstrate commercial 3D technology: VIP – for an ILC vertex detector, VICTR – for a CMS L1 "track-trigger" concept described in this section, and VIPIC - for a X-ray imaging correlation spectroscopy. Such a 3D-IC device can potentially provide O (10 ps) time resolution, track angle measurement, and a few micron hit position in a single layer. Still, challenges remain to be addressed before 3D-IC future applications in particle physics; among them are technology availability at a reasonable cost, deployment of complex sensors with minimal dead area, and the use of adapted analogue amplifier design to the most advanced processes offered by CMOS foundries [178-180].

## 8. Detector R&D for Dark Matter Searches, Astroparticle and Neutrino Physics

There are strong synergies in detector R&D studies between collider facilities and astroparticle, neutrino physics, and dark matter search experiments. Examples are the development of noble-liquid TPCs for a long-baseline neutrino program and dark matter searches and likely use of SiPMs in the future LAr (e.g. DarkSide, Argo, DUNE) and LXe-based detectors (e.g. DARWIN, nEXO). Dual-phase LAr TPCs, employing LEM/THGEM elements in the gas phase as a charge-readout option, proposed for the DUNE experiment; they are also being investigated in LBNO-DEMO at CERN [181]. In TPC-based detectors, both the scintillation (prompt) signal and the ionization (delayed) signal are readout, allowing for very powerful background discrimination. Careful studies of the details of the sensor medium response, for instance the nuclear recoil light yield as function of the drift electric field, are needed to properly calibrate the detector. Dual-phase TPCs use also the liquid phase for the WIMP conversion and an anode-separated gas phase for internal ionization charge multiplication. In the context of this paper, instrumentation trends in these domains are not systematically reviewed, but rather a few highlights are discussed, based on the presentations given at the INSTR2020 conference.

Ultra-high energy cosmic rays (UHECRs) are charged particles of energies above $5 \times 10^{19}$ eV that originate outside the Galaxy, a threshold known as the Greisen–Zatsepin–Kuzmin limit (GZK limit), assuming UHECR are protons. Recent progress in the field has been driven by the data collected by the Pierre Auger Observatory in the southern hemisphere and by the Telescope Array (TA) in the northern hemisphere, which cover areas of 3000 and 700 km$^2$ on the ground, respectively. The arrival directions of the highest energy cosmic rays are key to the understanding of the acceleration mechanisms at the far end of the spectrum. Because their flux at Earth is very small, the practical way of observing UHECRs is by measuring extensive air showers using simultaneously fluorescence devices (FD) and the giant array of surface detectors (SD) on the ground. As shown in Figure 17 (left), the SD is composed from 507 counters; each unit consists of 3 m$^2$ plastic scintillators read out by PMTs using wavelength-shifting fibers. The FD consists of three stations with a total of 38 telescopes overlooking the SD array. The two TA detector systems provide complementary information: SD measures the lateral distribution and time structure of shower particles arriving at the ground, and the FD measures the longitudinal development of the shower in the atmosphere and allows for a roughly calorimetric energy measurement (fluorescence light is proportional to the energy deposited along the shower path). Upgrade plans are underway to increase the area by a factor of 4 (TAx4) by adding 500 more scintillators. Such an enlarged area will be similar to that of the Auger Observatory; the array will be 95% effective above 57 EeV with an angular resolution of 2.2º. The main physics goals of the TA4 are better understanding of the energy spectrum and higher statistics in the TA Hot Spot region [182,183]. TAIGA (Tunka Advanced Instrument for cosmic ray physics and Gamma Astronomy) is a new hybrid detector system for ground-based gamma-ray astronomy for energies from a few TeV to several PeV, and for cosmic-ray studies from 100 TeV to several 100s of PeV. The key advantages of the future TAIGA installation is the joint operation of Tunka-HiSCORE, array of wide-WOV integrating Air Cherenkov stations, covering an area of initially ~1 km² and ~10 km² at a later phase of the experiment, and Tunka-IACT, consisting of 16 Imaging

Atmospheric Cherenkov Telescopes with a reflector area of 10 m² equipped with cameras of 400 PMT-based pixels [184]. The timing array Tunka-HiSCORE alone has a poor gamma-hadron separation at low energies (10–100 TeV), therefore, an underground muon detector (~3000 m²) is foreseen to determine the type of a primary particle [185,186].

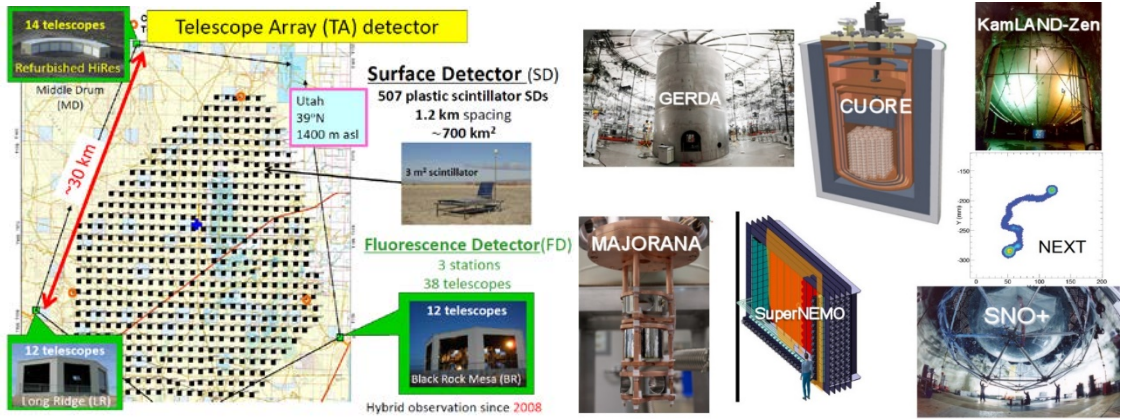

**Figure 17.** (left) The Telescope Array (TA) experiment with surface detector units represented by the dots and the FDs located on the perimeters of the sites; (right) Examples of neutrinoless double beta decay experiments.

Coherent elastic neutrino-nucleus scattering (CEνNS) plays the dominant role in the Universe in the processes where intense neutrino fluxes are generated; e.g. supernova bursts release ~ 99% of their energy to neutrino radiation. This process is predicted in the Standard Model and precise measurement of the cross-section is very important for astrophysics. Recently, COHERENT collaboration reported an experimental evidence (~3σ level) using the single-phase LAr detector at the Spallation Neutron Source at ORNL [187]. In parallel, the RED-100 collaboration will study CEνNS process using neutrinos from Kalinin nuclear power plant in the two-phase LXe detector readout by Hamamatsu PMTs [188].

Direct WIMP dark matter detectors aims to observe low energy (1-100 keV) nuclear recoils using different types of physical signals: heat (phonons in a crystal), scintillation light, or direct ionization of the target atoms. Several designs exploiting various target materials exist; in particular, LAr allows exquisite discrimination between nuclear and electron recoils via pulse-shape discrimination of the scintillation signals. Given the strong potential for the LAr technology to push the sensitivity for WIMP detection several orders of magnitude beyond the current levels, scientists from several groups (ArDM, DarkSide-50, DEAP-3600, and MiniCLEAN) have joined the Global Argon Dark Matter Collaboration (GADMC) to pursue a sequence of future experiments. The GADMC will require the use of large volumes of low-radioactivity argon depleted in $^{39}$Ar, necessitating the development of an underground argon extraction and purification plant. The immediate objective is the DarkSide-20k LAr two-phase detector, currently under construction at LNGS [189]. New type of SiPM suitable for LAr temperature operation was developed in collaboration with FBK. Studies of Ar-scintillation and electroluminescence properties have been reported in [190,191]. CEνNS effect represents an irreducible nuclear recoil background for dark matter searches, mostly for high-exposure detectors.

The T2K experiment is a long baseline neutrino oscillation experiment. To achieve CP violation measurement to better than 3σ confidence level until 2026, an upgrade is being planned for the beam power from 485 kW to 1.3 MW and for the current T2K Near Detector (ND280) complex [192]. The T2K-II will consist of a Super Fine Grain Detector (SuperFGD), acting as a neutrino target, with a 3D fiber tracker and SiPM readout [193], two high-angle Time Projection Chambers (HA-TPC) [194] and a TOF detector. New readout system is being developed for the HA-TPC, based on resistive MM concept, profiting from some synergies in R&D studies with ILC-TPC. T2K-II will be followed by the construction of a new far detector, HyperKamiokande (HK), which will profit from the development of novel MaPMTs with a larger sensitive area coverage per module with directional information.

The next generation of neutrinoless double-beta decay ($0\nu\beta\beta$) experiments is sensitive to the inverted mass hierarchy. Some examples of neutrinoless double-beta decays are shown in Figure 17 (right). Two types of experimental approaches have been commonly used to detect $0\nu\beta\beta$ signal: calorimetric, when the source is embedded in the detector, and the tracking-calo method. High energy resolution (up to FWHM=0.1% at $Q_{\beta\beta}$) can be achieved with Germanium (Majorana, GERDA, Legend) and bolometer detectors (CUORE, CUPID, AMoRE). However, it is difficult to reconstruct the event topology and to identify the background components, with the exception of the Xenon TPC (nEXO, EXO-200, DARWIN), but at the price of a lower energy resolution. Modern high pressure (10-15 bars) Xe-TPCs (NEXT, PandaX-III, AXEL) have two advantages over LXe TPCs: better energy resolution by proportional luminescence (demonstrated FWHM=1% at $Q_{\beta\beta}$, aim at 0.5%) and access to topological features, which may provide an extra discrimination from backgrounds. Recently, large existing liquid scintillator detectors, initially developed for neutrino oscillation measurements (Kamland, SNO), have been reused by adding an isotope inside them. It allows to reach quickly large amount of isotope (~100 kg) but with a limited energy resolution and thus non negligible background. Switchable fluorescent molecules to image tracks and measure energy in liquid $0\nu\beta\beta$ detectors beyond the tonne-scale represent new avenues, traditionally not exploited in HEP (e.g. Barium tagging) [195]. The tracking-calo method (Super-NEMO) separates detector from the source and allows to reconstruct directly the track of each of the two emitted electrons from the source foil and also to identify and measure each background component; the price is a lower efficiency and energy resolution. In general, the search for $0\nu\beta\beta$ requires several experimental techniques and more than one isotope. Nuclear matrix elements are not always well known and there might be unknown background and gamma transitions at the end point of a particular isotope [195,196].

## 9. Summary and Outlook

Over the course of the last fifty years, the progress in experimental particle physics was driven by the advances and breakthrough in instrumentation, leading to the development of new, cutting-edge technologies and providing the platform for the training of scientists and engineers of tomorrow. A close interplay between generic R&D, future project-oriented studies and the R&D collaborations will be needed in the future to ensure an efficient strategic alignment and a coherence in the R&D activities, whilst preserving accessibility of results and know-how to the wider community. There is a need to strengthen the relations between academia and industry in the field of scientific instrumentation. The development of modern large-scale detector systems presents numerous challenges, which are at the same time of the major technological, engineering and organizational complexity. A constant investment into generic "blue sky" R&D is mandatory to retain the technical and engineering expertise to address challenges of the future colliders.

Timely R&D studies across many communities at the Energy/Intensity/Cosmic Frontiers is expected to enhance developments of common technological platforms. High priority must be given to the education in instrumentation of the next generation of young researchers, to be able to cope with the future detector challenges. In addition, excellence in instrumentation development is not recognized enough at the universities to foster the participation of young people in this branch of science. Special programs which allow equal-opportunity careers in physics and instrumentation at the universities, for both students and professors, could further enhance the breadth and depth of the high energy physics domain in the future.

**Acknowledgments**